\documentclass[twocolumn,floatfix]{aastex63}
\usepackage{color}
\usepackage{listings}
\usepackage{url}
\usepackage{verbatim}
\usepackage{subfigure}
\usepackage{amsmath}
\usepackage{hyperref}
\usepackage{booktabs}

\definecolor{dkgreen}{rgb}{0,0.6,0}
\definecolor{gray}{rgb}{0.5,0.5,0.5}
\definecolor{mauve}{rgb}{0.58,0,0.82}
\begin{document}

\title{The Dependence of the Type Ia Supernova Host Bias on Observation or Fitting Technique}
\shorttitle{SN Ia Host Bias Dependence on Technique}
\shortauthors{Hand et al. 2021}

\correspondingauthor{Jared Hand}
\email{jsh89@pitt.edu}

\author[0000-0001-7260-4274]{Jared Hand}
\affiliation{
    Pittsburgh Particle Physics, Astrophysics, and Cosmology Center (PITT PACC).
    Department of Physics and Astronomy,
    University of Pittsburgh,
    Pittsburgh, PA 15260, USA
}
    
\author{Shu Liu}
\affiliation{
    Pittsburgh Particle Physics, Astrophysics, and Cosmology Center (PITT PACC).
    Department of Physics and Astronomy,
    University of Pittsburgh,
    Pittsburgh, PA 15260, USA
}

\author[0000-0002-1296-6887]{Llu\'is Galbany}
\affiliation{
    Institute of Space Sciences (ICE, CSIC), Campus UAB, Carrer de Can Magrans, s/n, E-08193 Barcelona, Spain.
}

\author[0000-0002-3988-4881]{Daniel Perrefort}
\affiliation{
    Pittsburgh Particle Physics, Astrophysics, and Cosmology Center (PITT PACC).
    Department of Physics and Astronomy,
    University of Pittsburgh,
    Pittsburgh, PA 15260, USA
}
    
\author[0000-0001-7113-1233]{W.~M. Wood-Vasey}
\affiliation{
    Pittsburgh Particle Physics, Astrophysics, and Cosmology Center (PITT PACC).
    Department of Physics and Astronomy,
    University of Pittsburgh,
    Pittsburgh, PA 15260, USA
}
\author[0000-0003-4625-6629]{Chris Burns}
\affiliation{
    Observatories  of  the  Carnegie  Institution  for  Science,  
    813 Santa Barbara St,
    Pasadena, CA, 91101, USA
}

\begin{abstract}
    More luminous Type Ia supernovae (SNe~Ia) prefer less massive hosts and regions of higher star formation. 
    This correlation is inverted during width-color-luminosity light curve standardization resulting in step-like biases of distance measurements with respect to host properties.
    Using the PISCO supernova host sample and SDSS, GALEX, and 2MASS photometry, we compare host stellar mass and specific star formation rate (sSFR) from different observation methods, including local vs. global, and fitting techniques to measure their impact on the host step biases.
    Mass step measurements for all our mass samples are consistent within a 1$\sigma$ significance from -0.03$\pm$0.02~mag to -0.04$\pm$0.02~mag. 
    Including or excluding UV information had no effect on measured mass step size or location.
    Specific SFR (sSFR) step sizes are more significant than mass step measurements and varied from $0.05\pm0.03$~mag (H$\alpha$) and $0.06\pm0.02$~mag (UV) for a 51 host sample. 
    The sSFR step location is influenced by mass sample used to normalize star formation and by sSFR tracer choice. 
    The step size is reduced to 0.04$\pm$0.03~mag when using all available 73 hosts with H$\alpha$ measurements.
    This 73 PISCO host subsample overall lacked a clear step signal, but here we are searching for whether different choices of mass or sSFR estimation can create a step signal.  We find no evidence that different observation or fitting techniques choice can create a distance measurement step in either mass or sSFR. 
\end{abstract}

\section{Introduction} \label{sec:introduction}

The discovery of dark energy using Type Ia Supernovae (SNe~Ia) nearly 20 years ago established SN cosmology as a key field in observational cosmology \citep{Riess98, Perlmutter99}. SNe~Ia are often used as standardizable candles where empirical relationships are used to further reduce the already low observed intrinsic scatter at peak luminosity. In particular, the width-luminosity relationship accounts for a correlation in peak luminosity with SN~Ia duration (referred to as width or stretch) while the color-luminosity relationship accounts for a correlation between peak luminosity and color \citep{Phillips93, Riess96}. By applying these corrections, recent analyses have been able to reduce scatter to within $\sigma < 0.1$ mag  \citep{Conley11, Betoule14, Scolnic18, Brout2019}, achieving a level of precision comparable other cosmology probes such as CMB polarization, CMB temperature power spectrum decomposition, and lensing \citep{Aghanim2018}.  

Given the growing tension between $H_0$ measurements by competing probes \citep{Freedman2017}, high precedence is placed on further improving measurement precision.
Considering the rate and brightness of SNe~Ia, coupled with the expected
detection rate of next-generation facilities such as the Vera C. Rubin Observatory \citep{Zeljko2019} and Nancy Grace Roman Space Telescope \citep{Spergel2015}, the utility of these transient events to constrain the properties of dark energy remains high. Multiple concerted efforts to increase the number of SN~Ia observations across ever larger redshift and wavelength ranges have already reduced random error to the point that systematic errors have become the dominant source of uncertainty \citep{Wood-Vasey07, Kessler09, Conley11, Betoule14, Scolnic18, Brout2019}.
It is paramount to understand and reduce these uncertainties in order to maximize the utility of upcoming surveys.

Brighter (dimmer) SNe~Ia prefer less (more) massive and star-forming (quiescent) host galaxies --- this trend is reversed post-standardization.  The most commonly used host property to quantify this bias is stellar mass (see Section \ref{sec:background} for details), with recent SN~Ia analyses incorporating a luminosity correction term for host mass during standardization \citep{Conley11, Betoule14, Scolnic18}. This effect is typically quantified against Hubble residuals $\mu_{\rm SN} - \mu_{\rm mod}(z_{\rm SN})$, or the difference between the standardized brightness and that predicted by the best-fit cosmology. The ubiquitous host bias measured post-standardization is the mass step, but similar step-like trends between specific star formation rate (sSFR) and Hubble residuals for both global and local 1~kpc apertures have been observed \citep{D_Andrea11, Rigault18}.  As dust, age, and metallicity correlate with mass and sSFR, similar trends have been observed for they properties; all trends are more pronounced when compared to stretch (see \ref{sec:background} for more details).
The existence of these biases in historical data sets is not ubiquitous, though. Variation in trends between SFR and SN~Ia properties based on SFR calibration was observed in \cite{D_Andrea11}, and no mass step was found in the initial analysis of DES SNe~Ia \citep{Brout2019}; subsequent DES analyses have measured a mass step \citep{Smith2020, Kelsey2021}.  This prompts the question: does the measured host-property bias depend on the chosen observation method and fitting technique used to estimate host galaxy properties?

Any attempt to explain the relationship between SNe~Ia and their host galaxy's properties should begin with quantifying whether the significance of observed trends vary between choice of observation method and fitting technique. The host bias extends from the local 1~kpc environment to the entirety to a global aperture; integral field spectroscopy (IFS) samples such as PISCO \citep{Galbany2018}, in providing resolved spectral information of a galaxy, are an ideal tool to measure the host bias.  Pragmatism has necessitated the use of photometry to constrain host properties in nearly all cosmology analyses. We considered the following questions:
\begin{enumerate}
    \item Does observational technique have a significant effect on measured global and local host bias?
    \item Does the fitting technique used to estimate host properties effect the measured host bias?
\end{enumerate} 

We here built a set of of host galaxy samples and then inferred galaxy properties using different observational and fitting techniques.
We compared stellar mass and SFR estimates from the PISCO sample.
Using the PISCO SN~Ia subsample, we then compared different stellar mass and sSFR estimate samples to SN~Ia properties, looking primarily for differences in the resulting host-SN~Ia property trends between mass/sSFR estimate samples.  Finally, the larger observed size and strength of the sSFR host bias relationship relative to the stellar mass host bias is both presented and discussed.

In Section~\ref{sec:background} we summarize SN~Ia host studies up to the present followed by a summary of the considered data set in Section~\ref{sec:data}. Section~\ref{sec:methodology} provides an overview of our methodology, including summaries of fitting techniques used. Our results are provided and discussed in Section~\ref{sec:analysis}, and concluding remarks presented in Section~\ref{sec:conclusion}.
\section{Summary of Host Galaxy Bias Literature} \label{sec:background}

The presence of an observed correlation between the properties of SNe~Ia and their host's properties is common in the literature.  
Parameters of light curve width variation (such as $x_1$ in SALT2 \cite{Guy2007}) see lower $x_1$ values prefer more massive hosts, older stellar populations, and environments of larger SFR \citep{Sullivan10, Kelly10, Gupta11, Conley11, Betoule14, Scolnic18}.  Both \cite{Sullivan10} and \cite{Gupta11} saw a relationship between the SALT2 color parameter $c$ and host mass and host age, respectively, albeit at a lower significance than $x_1$; \cite{Scolnic18} found no significant color trend.  Recently, \cite{Pruzhinskaya2020} found a statistically significant dependence of SALT2 $x_1$ on host galaxy morphology, with lower $x_1$ SNe Ia preferring elliptical and lenticular galaxies. \par

The observed correlation between SNe~Ia and host properties are not explicitly included nor accounted for with a canonical Tripp standardization model \citep{Tripp1998}, resulting in an apparent over-correction of the host bias post-standardization.
For example, both the SN Legacy Survey \citep[SNLS;][]{Sullivan10} and SDSS SN Survey \citep{Sako18} found a `mass step', where SNe~Ia with brighter Hubble residuals were on average hosted by more massive galaxies \citep{Kelly10, Lampeitl10, Gupta11}.   
In comparing residuals in the fitted distance modulus of SNe Ia with respect to cosmological prediction to their host masses, both SDSS and SNLS samples observed a sudden step-like change in Hubble residual vs mass near $10^{10} M_{\odot}$. 
\cite{Conley11}, \cite{Betoule14}, and \cite{Scolnic18} also observed this mass step at $10^{10} M_{\odot}$ in their combined SN Ia sample analyses. 
Initial analysis by \cite{Brout2019} of Dark Energy Survey (DES) SN Ia found no evidence for a mass step post-standardization, although more thorough methodology for an updated SN~Ia sample did indeed recover a mass step \citep{Smith2020, Kelsey2021}.  
Using SNIFS spectrophotometry from SNFactory, \cite{Rigault13} and \cite{Rigault18} found that Hubble residuals are strongly dependent upon SFR and sSFR, respectively, the latter having observed a $5\sigma$ relationship between Hubble residual and sSFR and confirming the tentative findings of \cite{D_Andrea11}.  Recently, \cite{Jones18} compared $u-g$ color of SN Ia 1.5 kpc local environment apertures against randomly selected 1.5 kpc apertures within the host, finding that the SN Ia environment correlates with SN Ia distance slightly more than random apertures, and found a local stellar mass step after the global mass step had been corrected for.  \cite{Hayden2013} used both mass and metallicity to reduce Hubble residual scatter.  \cite{Roman2018} used $U-V$ as a proxy for stellar age and found an age mass step of similar magnitude and significance as the fiducial mass step, and found a $7\sigma$ relationships between local $U-V$ measurement and distance.  \cite{Rose2019}, using principal component analysis, found a strong relationship between a linear combination of host galaxy properties and Hubble residuals. 
\cite{Brout2020} presented a model that explain the mass step post-light curve standardization via SALT2 resulting from improper treatment of host dust, which was further expanded by \cite{Popovic2021}. In contrast, \cite{Uddin2020} found a clear dependence on Hubble residuals calculated from light curves fit with SNooPy with host mass calculated for visible and NIR light curves that was inconsistent with the model presented by \cite{Brout2020}.  Both \cite{Gonzalez-Gaitan2020} and \cite{Ponder2020} found tentative evidence for a host mass step for SNooPy-fit NIR light curves for SN~Ia.

The use of simple stellar population (SSP) templates and stellar population synthesis (SPS) libraries has been a central part any endeavor to estimate a galaxy properties; these properties include stellar mass \citep{Courteau2014}, stellar age, star formation rate \citep{Kennicutt1998, Kennicutt2012}, and gas-phase/stellar metallicity \citep{Tremonti2004, Gallazzi2005, Kewley2008}. Likewise, complex systematic uncertainties in modeling and from observational limitations propagate into uncertainties in derived property estimates.  Past SN Ia cosmology analyses have generally treated SSP/SPS libraries and fitting software as black boxes when estimating host properties, and frequently relied on few data points from photometry spanning a narrow wavelength range. Redshift, stellar mass, stellar age, dust extinction, metallicity, and star formation rate are often degenerate with each other and large systematic covariances complicate fitting methods \citep{Kauffmann2003, Brinchmann2004, Tremonti2004, Gallazzi2005}.
Such degeneracies, together with systematic uncertainties, result in dramatic uncertainties for some best fit parameters (see section 3 introduction for more information). \par

Although optical multi-wavelength photometry of host galaxies has been the primary tool used to estimate host mass and SFR \citep{Sullivan10, Kelly10, childress2013, Betoule14, Scolnic18, Brout2019}, optical wavelength spectra \citep{D_Andrea11, Johansson2013}, UV or IR photometry in addition to optical spectra or photometry \citep{Gupta11, Anderson2015}, and IFS \citep{Rigault13, Rigault18, Galbany2018} have been used.
\cite{childress2013} and \cite{Gupta11} both found stellar mass estimates using optical, IR, and UV photometry changes little from mass estimates using optical photometry only, consistent with the results of \cite{Bell&deJong2001}.
\cite{D_Andrea11} used sSFR estimates from SDSS~II spectra and two different mass estimates: one from SPS model best fits using SDSS~II spectra and one from \cite{Gupta11}.
\cite{Hayden2013} directly compared three different mass estimates for a set of SDSS host galaxies using metallicity estimates from the Fundamental Metallicity Relation \citep{Mannucci2010} to further reduce Hubble residual scatter. \par 

\section{Data} \label{sec:data}

We consider 319 galaxies observed as part of the PMAS/PPak Integral-field Supernova Hosts Compilation \citep[PISCO][]{Galbany2018}.

\begin{figure}
    \centering
    \includegraphics[width=\linewidth]{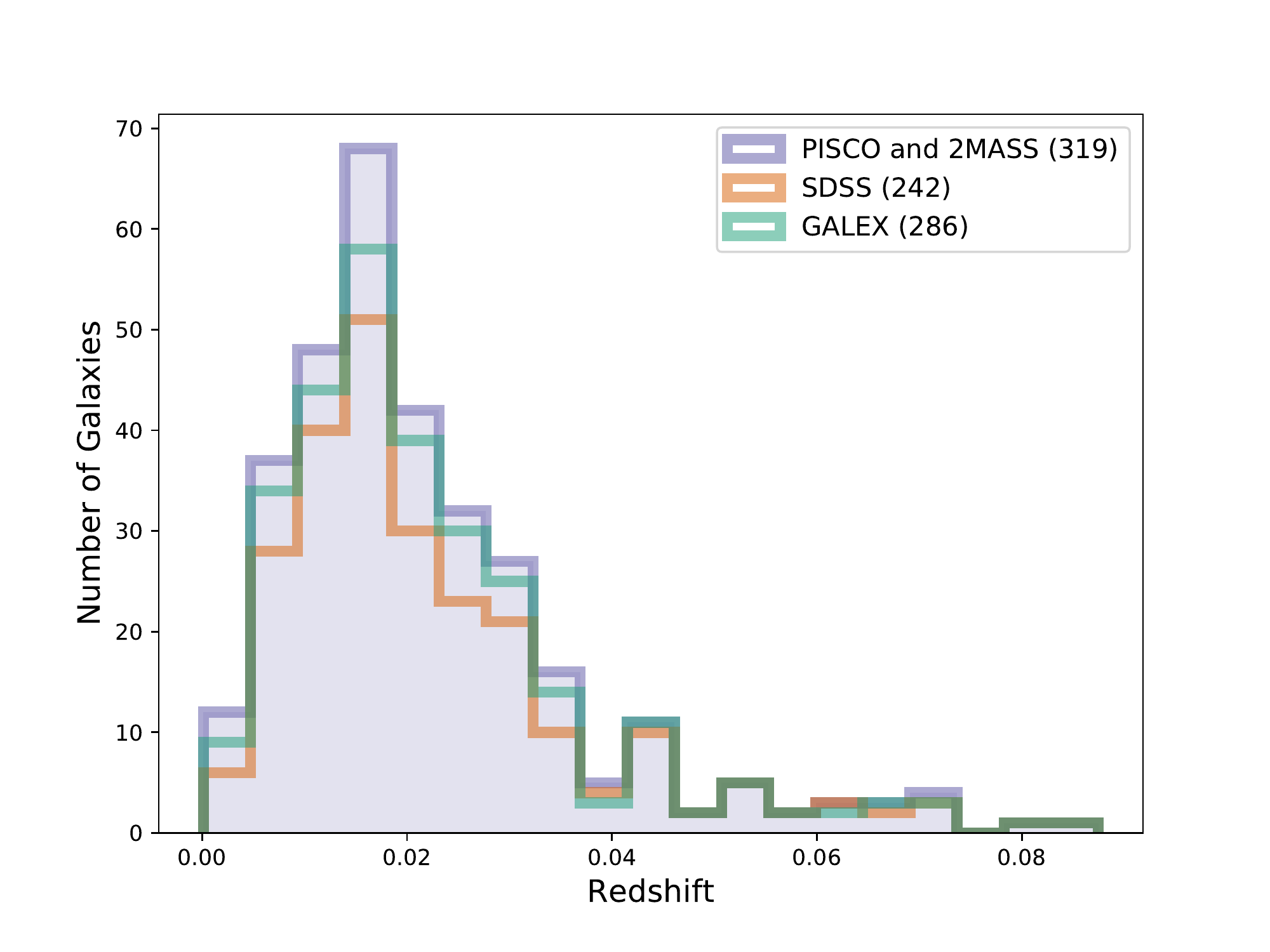}
    \caption{Distribution of redshifts for galaxies considered by this paper. This includes a total of 319 galaxies selected from the PISCO sample with supplementary observations from 2MASS, SDSS DR12, and GALEX. All considered targets were observed by PISCO and detected in 2MASS.}
    \label{fig:my_label}
\end{figure}

These observations were complemented by optical \textit{ugriz} photometry from the Sloan Digital Sky Survey (SDSS), near-infrared (NIR) \textit{JHK$_S$} photometry from the Two Micron All Sky Survey (2MASS), and ultraviolet (UV) NUV+ and FUV photometry from the Galaxy Evolution Explorer (GALEX). 
Figure~\ref{fig:my_label} shows the redshift distribution of the sample, which spans a range of $0.00013 < z < 0.0875$.
 These 319 galaxies host 375 SNe,
of which 198 are SNe~Ia.

PanSTARRS DR1 \citep{Chambers2016} was considered to expand coverage, but added complexity with overlapping visible wavelength photometry and our constraint to northern hemisphere targets resulted in its disregarding in favor of SDSS.  Follow up analyses will likely make use of PanSTARRs DR2, especially if our observation footprint extends into the southern hemisphere.

\subsection{PISCO Supernova Sample} \label{sec:data:pisco_sn}

PISCO is a compilation of SN host galaxies observed with integral field unit (IFU) spectroscopy using the same instrument PMAS/PPak \citep{2005PASP..117..620R,2006PASP..118..129K} mounted to the 3.5~m Calar Alto telescope.
PISCO was built from an initial sample of SN host galaxies obtained by the CALIFA survey \citep{2016A&A...594A..36S,Galbany2014,2016A&A...591A..48G} and extended with new observations of SN hots based on different science goals (see \citealt{Galbany2018} for details).

The PISCO SN~Ia subsample had 198 objects.
When calculating distance moduli all 14 peculiar SNe~Ia were ignored (e.g., SN~1991bg and 1991T-like objects), while 83 normal SNe~Ia lacked available optical photometry, with many only having NIR light curves.  SN~2014J was excluded due to its extremely low redshift and host galaxy M82's mass estimation difficulties.
We used the SALT2 output fits for 30 unpublished SNe~Ia from Carnegie Supernova Project II (CSPII, Suntzeff et al, in prep), bringing the sample to 101.
During SALT2 fitting we removed SN~2006lf, which had an very high Milky Way color excess of $E(B-V)=0.821$, giving a final total of 100 SNe~Ia used.  SDSS coverage reduced this count to 76, while mutual SDSS and GALEX coverage further reduced the number to 66 (Section \ref{sec:analysis:host_mass_bias_comp}).  Four SN~Ia hosts lacked H$\alpha$~flux measurements and 11 host galaxies had unavailable GALEX FUV or NUV~flux measurements.  These too were all excluded, bringing the total SN Ia count to 51 for sSFR analysis (Section \ref{sec:analysis:ssfr_bias_comp}). See Tables \ref{table:snia_cuts} and \ref{table:snia_subsamples} for a summary of sample reductions.

\begin{deluxetable}{lr}
\tablecaption{Summary of SNe Ia sample size reductions. Details on specific SN Ia removal is provided in Section \ref{sec:data:pisco_sn}  Each row defines a subgroup of the row described above.} \label{table:snia_cuts}
\tablehead{\colhead{SN Ia Subsample Condition} & \colhead{SN Ia Count}}
\startdata
Total PISCO SNe Ia & 198 \\
Normal SNe Ia & 184 \\
Available visible photometry + CSPII & 101\\
Excluding specific SNe Ia & 100\\
Available STARLIGHT mass estimates & 97\\
\enddata
\end{deluxetable}

\begin{deluxetable}{lr}
\tablecaption{Summary of mass and sSFR step analyses subsample sizes starting with 97 available STARLIGHT mass estimates.} \label{table:snia_subsamples}
\tablehead{\colhead{SN Ia Subsample} & \colhead{SN Ia Count}}
\startdata
SDSS coverage & 76 \\
SDSS+GALEX coverage & 66 \\
Available H$\alpha$ measurements & 73 \\
Available H$\alpha$ and SDSS+GALEX coverage & 51\\
\enddata
\end{deluxetable}

\subsection{PISCO Host Galaxy Global Parameters} \label{sec:data:pisco_gal}

Stellar mass estimates and integrated H$\alpha$ flux measurements of all PISCO galaxies were taken from \cite{Galbany2018} with mass estimates made using STARLIGHT \citep{CidFernandes2005}. Mass estimates were calculated using a SSP library built using \cite{2003MNRAS.344.1000B} (BC03) with a \cite{2003PASP..115..763C} initial mass function (IMF).
\cite{Galbany2018} used a more complex set of basis templates  instead of just BC03 used here (see \cite{Galbany2018}, Section 3.2.1 for more details) --- otherwise, the methodology we used to estimate masses with STARLIGHT was identical.

\subsection{SDSS DR12 Optical Photometry} \label{sec:data:sdss}
The Sloan Digital Sky Survey (SDSS) is a multi-year, multi-program survey performed at the Apache Point Observatory (APO) using the SDSS 2.5~m telescope \citep{York00, Gunn06}.
We used observations from SDSS Data Release 12 \citep[DR12][]{Alam2018} to provide imagining data for host galaxies in the \textit{ugriz} filters \cite{Doi10}. 
Mutual coverage between SDSS and PISCO is available for 239 of the 319 galaxies (74.9\%). 

All SDSS fields were selected using the nearest neighbor search functions built into the DR12 CASJOBS database\footnote{\url{http://skyserver.sdss.org/CasJobs/SchemaBrowser.aspx}} and downloaded from the Science Archive Server (SAS)\footnote{\url{https://dr12.sdss.org/}}.
 
\subsection{2MASS IR Photometry} \label{sec:data:2mass}
The Two Micron All Sky Survey (2MASS) was a three-year program run at the Whipple Observatory and Cerro Tololo Inter-American Observatory \citep{Skrutskie2006}. 2MASS observed 99.998\% of the sky with three near infrared (NIR) passbands \textit{JHK$_s$}, resulting in all PISCO galaxies having corresponding 2MASS images. Optimal operations result in a 2.5--3.4\arcsec PSF with a Pixel size of $2.0$\arcsec.
Images were obtained using the Infrared Science Archive (IRSA).\footnote{\url{https://irsa.ipac.caltech.edu/frontpage/}}

\subsection{GALEX Photometry} \label{sec:data:galex}
The Galaxy Evolution Explorer (GALEX) is a space-based observatory with a 1.2~degree diameter circular field of view that observed through two ultra-violet (UV) bands,
FUV (135--175~nm) and NUV (175--280~nm), 
\citep{Martin2005}.
A total of 287 PISCO galaxies reside within GALEX's footprint for a mutual coverage of 90.0\%, although only 203 were also within SDSS coverage.
Images were downloaded from the four primary GALEX surveys using STSci MAST services \footnote{\url{https://mast.stsci.edu/portal/Mashup/Clients/Mast/Portal.html}}, with precedent placed on exposure time when choosing fields.
Where possible we used ``Deep Imaging Survey'' fields (DIS: 30,000~s exposure), followed by ``Medium Imaging Survey'' fields (MIS: 1500~s exposure), and finally the ``All-sky Imaging Survey'' fields (AIS: 100$\times$10~s exposures \cite{Martin2005}.
If available, Nearby Galaxy Survey (NGS) fields were used instead of AIS fields.
Guest investigator fields were also used when necessary.
2.2\% of PISCO galaxies had DIS coverage, 5.1\% had NGS coverage, 14.4\% had MIS coverage, and 80.9\% had AIS coverage or were observed as part of the guest investigator program.

\section{Methodology} \label{sec:methodology}
We performed our global photometry and combined that with corresponding PISCO IFS to estimate host galaxy properties.
For details on PISCO data reduction we refer the reader to  \cite{Galbany2018}.
As one brief note, the PISCO IFS apertures were hexagonal, while the apertures on the imaging data we analyzed here are elliptical.

\subsection{Photometric Image Preparation} \label{sec:methodology:photometry}
Foreground objects in images from each data set were removed by masking. 
The position and appropriate aperture size of foreground objects in SDSS images were determined using Source Extractor \citep{Bertin1996} on \textit{r}-band images.
All identified objects were visually reviewed and assigned a proportion factor ($\alpha_{\rm SDSS}$) which corrected the Kron radius calculated by Source Extractor ($r_k$)

\begin{equation}
    r_k^{\rm cor} =  \alpha_{\rm SDSS}~r_k
\end{equation}

This factor was used to decrease the area of measured foreground objects so to not compromise the shape of the host galaxy.
Pixel values within the scaled aperture were masked and replaced using an interpolation to account for lost photons in the masked region. 

For SDSS, we replaced all foreground objects within a circular region of the host less than $A \times r_k^{\rm cor}$ using a two-dimensional linear interpolation. With GALEX and 2MASS, we instead masked foreground objects due to lower field signal-to-noise leading to unrealistic interpolation results. 
All foreground objects outside this radius but within a radius of $2.8 (A \times r_k^{\rm cor})$ were simply masked and unused. 
For 2MASS and GALEX we masked all foreground objects within a radius of $2.8 (A \times r_k^{\rm cor})$. SDSS images were provided background-subtracted with FITS~header information for background reconstruction. Background images were provided with GALEX field downloads and were used directly in background subtraction.  Zero pixels in both image and provided background fields were masked for GALEX NUV and FUV data sets.  As both fields were masked, this did not increase our background estimation.  

\subsection{Galaxy Parameter Estimation} \label{sec:methodology:gal_param_est}
To explore how the host galaxy bias varies with fitting technique, we required host property estimates from multiple techniques. To this end, we used three fitting programs to estimate host properties two for photometry and one for IFS.  These fitting techniques make use of simple stellar population (SSPs) libraries, or use stellar population synthesis (SPS) libraries built by convolving a set of SSPs with a star formation history (SFH) parameterization and a recipe to handle dust extinctions to produce a library of composite stellar populations (CSPs) \citep{Walcher2011, Conroy2013}.  It is important to explicitly note that simple stellar population (SSP) or stellar population synthesis (SPS) libraries contain numerous sources of systematic uncertainties  and that even with high-quality observations, systematic model uncertainties endemic to SPS libraries are larger than what is expected from observational errors alone \citep{Conroy2009}. 

\begin{figure}
    \centering
    \includegraphics[width=\linewidth]{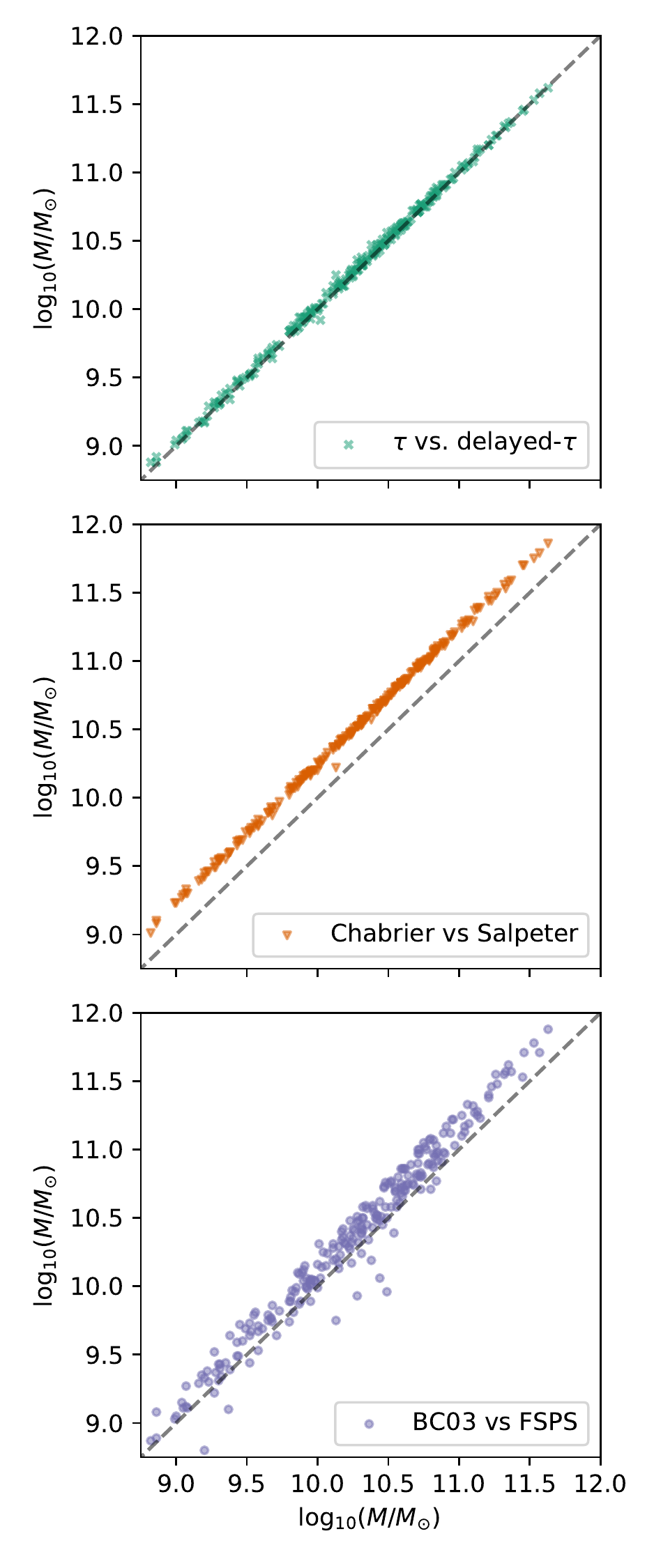}
    \caption{Systematic variation in estimated mass (left column) from variation in SFH (top), IMF (middle), and SPS library (bottom).
    Along the x-axes are the base template sets with a delayed-$\tau$ SFH (top), a Chabrier IMF (middle), and a BC03 library (bottom).
    Along the y-axes are
    (top) an exponential decay $\tau$ model,
    (middle) a Salpeter IMF,
    (bottom) FSPS.}
    \label{fig:sps_sys}
\end{figure}

\subsubsection{ZPEG} \label{sec:methodology:gal_param_est:zpeg}
ZPEG estimates galaxy redshifts by fitting photometry against a library of CSPs constructed from the theoretical spectral library PEGASE.2. The best-fit template is determined using maximum likelihood analysis with a $\chi^2$ function for the difference between synthetic photometry computed from templates and observed photometry.  Specifically, $\chi^2$ values are calculated for each point along a four-dimensional grid of parameters ($E(B-V)$, $t_*$, redshift, and template), with the best-fit model being the minimal $\chi^2$ value on the grid
\begin{equation}
\chi^2_Z = \sum^{N}_i \frac{\big[F_{\lambda, i} - \alpha F_{\lambda, \text{mod}}\big(t_*,E(B-V),z\big)\big]^2}{\sigma^2_i}
\end{equation}
where $N$ is the number of filters, $F_\text{mod}$ is the model flux density from given template of age $t_*$ and color excess $E(B-V)$, $\sigma^2_i$ is the measured observation variance for data point $i$, and $\alpha$ is a scaling parameter to match normalized template SEDs to the observed flux. Uncertainty in the best-fit model is determined by calculating the reduced $\chi^2$ ($\chi^2_r$) value of the best-fit model and finding the corresponding model parameters whose model  $\chi^2$ ($\chi^2_{\rm err}$) satisfy $\chi^2_{\rm err}\leq \chi^2_r + 1$. Spectroscopic redshifts can be provided for each object to simplify fitting.

With ZPEG, we used the PEGASE.2 library, a Salpeter IMF, included nebular emissions, and averaged SFR over 500 million years starting at present.
We used the default 200-ages Salpeter library of 15 galaxy templates provided with ZPEG, excluding the starburst template due to its extremely nonphysical description of galaxy evolution.
A foreground dust screen with color excess $E(B-V)$ with values ranging from $0$ to $0.4$ mags in intervals of $0.2$ mags was added for each template when fitting.
There is a setting ``AGE\_CONSTR\_Z0'' for ZPEG that when turned off can result in dramatic underestimation of mass, especially if the starburst template is included.  We kept the default setting of ``T'' to keep it enabled.

\subsubsection{FAST++} \label{sec:methodology:gal_param_est:fast++}
FAST++ is a C++ implementation of FAST \citep{Kriek2009} with added functionality \citep{FAST++, Schreiber2018}.  Given photometry or spectra, FAST++ determines a best-fit SED from a provided SPS library on a five-dimensional grid, with each grid point corresponding to a CSP SED with age $t_*$, SFH time scale $\tau$, $V$ band extinction $A_V$, and stellar metallicity $Z_*$ at some redshift $z$.  Note that the CSP  is constructed using the selected SFH parameterization and dust extinction model.  The redshift parameter can be fixed by providing a spectroscopic redshift value as input for fitting.  At each point on the grid a $\chi^2$ value is calculated:
\begin{equation}
    \chi^2_F = \sum^{N}_i \frac{\big[F_{\lambda, i} - F_{\lambda, \text{mod}}(t_*, \tau, A_V, Z_*, z)\big]^2}{\sigma^2_i}
\end{equation}
where $N$ is the number of combined photometric and spectroscopic data points and $\sigma^2_i$ is the observation variance for data point $i$.  Confidence intervals for parameters and derived quantities (such as stellar mass and SFR) are determined using Monte Carlo sampling of the grid around the lowest $\chi^2$.  

SFR was averaged over the last 500 million years as done with ZPEG.  Unless otherwise stated, we used a Calzetti dust law with a uniform and constant foreground dust screen with the BC03 SPS library calculated using a delayed exponential (delayed-$\tau$) SFH and a Chabrier IMF \citep{Calzetti2000, Chabrier2003}.

Using FAST++ we compared the systematic effects of changing different SPS library components (Figure~\ref{fig:sps_sys}).
Mass estimates varied uniformly with variation in IMF (Chabrier vs. Salpeter) and were unchanged for different SFH models (exponential or $\tau$ vs. delayed-$\tau$).  The global shift from changing to a Salpeter IMF was 0.24~dex. Leaving IMF and SFH fixed, when then compared mass estimates for the SPS libraries from BC03 to the newer FSPS \citep{Conroy2009}.  The particular FSPS library used was the default provided by FAST/FAST++ as described in \cite{Aird2017}, Appendix A. 
FSPS estimated systematically higher mass at $0.12$~dex with a standard deviation of $0.12$~dex --- a higher scatter than that due to changing IMF or SFH but well within the average error in mass estimates and the expected 0.3~dex template uncertainty.

\subsubsection{Fitting IFS with STARLIGHT}
\label{sec:methodology:methodology:gal_param_est:inversion_Fitting}

STARLIGHT fits observations to linear combinations of SSPs attenuated by a given dust law in a process called spectral inversion fitting \citep{CidFernandes2005}.  This differs from FAST++ and ZPEG which instead fit observations to a library of composite stellar populations, themselves built from SSPs using some star formation history and dust law. 

STARLIGHT fits spectra by masking emission lines.  A Fitzpatrick dust model with $R_V=3.1$ was used for dust correction \citep{fitzpatrick1999}. The best-fit SED is determined by minimizing the resulting $\chi^2$ value against the model SED $M_{\lambda}$ using an implementation of the Metropolis algorithm \citep{CidFernandes2005}, with:
\begin{equation}
    M_{\lambda} = M_{\lambda, 0} 10^{-0.4(A_{\lambda} - A_{\lambda,0})} \Bigg[\sum^{N_\ast}_{i}x_i F_{\lambda,i}\Bigg] \odot \mathcal{N}(v_\ast, \sigma_{v_\ast})
\end{equation}
where $x_i$ and $F_{\lambda,i}$ are the fractional contribution and synthetic flux density of $i$th SSP SED, respectively, $\odot$ is the convolution operator, and the normal distribution $\mathcal{N}$ models velocity dispersion due to stellar motion $v_*$.  $M_{\lambda,0}$ is the normalized synthetic flux density.  The population vector elements $x_i$ serve as weights for each $F_{\lambda,i}$, from which one can infer SFH, metallicity, and other properties as a weighted sum of the properties of each contributing $F_{\lambda,i}$. To improve consistency with FAST++'s BC03 and the Chabrier IMF template library described above, the PISCO mass estimates used here were recalculated using a SPS library constructed from the BC03 library and Chabrier IMF as well.  

\subsection{Hubble Residuals}
\label{sec:methodology:hres}

Hubble residuals are the difference between the measured distance modulus and that predicted by the assumed cosmology at a given redshift $z_{\text{CMB}}$: $\mu - \mu_{\text{mod}}(z_{\text{CMB}}; H_0, \Omega_{\Lambda})$.  
To better facilitate comparison with existing literature, Hubble residuals were estimated in a similar fashion to the Joint Light-Curve Analysis, or JLA \citep{Betoule14} using the light curve fitter SALT2. 

70 of 100 used PISCO SNe Ia had light curves fit with the SNCOSMO implementation of SALT2 using the \cite{Betoule14} fiducial light curve template \citep{Guy2007, Barbary2014}. In particular, we fit for the scaling parameter ($x_0$), the effective stretch parameter ($x_1$), and the color parameter ($c$). Milky Way dust extinction was taken into account during fitting using \cite{Schlegel1998} dust maps with a Cardelli dust law \cite{Cardelli1989}.  If a light curve was observed by more than one survey, then each set of observations were fit separately. The light curve fits were then visually inspected and the better fit was used.

The 30 remaining SNe Ia were observed as part of the unpublished CSPII (Suntzeff et al, in prep).  Preliminary SALT2 parameters were provided by CSPII for this analysis, as many low-mass PISCO SN Ia hosts consisted primarily of CSPII hosts.

To standardize fitted SN Ia luminosity, cosmological parameters were held fixed assuming a flat $\Lambda$CDM cosmology with $H_0=70$ and $\Omega_{\Lambda}=0.7$ using astropy's cosmology suite \citep{astropy13}. The distance modulus per SN is calculated assuming a linear relationship between both SALT2 color parameter $c$ and SALT2 stretch parameter $x_1$ versus absolute $B$-band magnitude $M_B$ with slopes $\alpha$ and $\beta$, respectively \citep{Tripp1998}:
\begin{equation}
    \mu_B = m_B + \alpha x_1 - \beta c - M_B
\end{equation}
Here, $\alpha$ and $\beta$ are nuisance variables calculated using a maximum likelihood analysis akin to \cite{Conley11}:
\begin{equation} \label{eq:standardization}
    \chi^2_H = \sum^{N_{\text{SN}}}_{i} \frac{\big( \mu_{B,i} - \mu_{\rm mod}(z_{\text{CMB},i}; H_0, \Omega_{\Lambda})\big)^2}{\sigma^2_i + \sigma^2_{\text{int}}}
\end{equation}

We ignored covariance between SNe for simplicity and fixed  intrinsic dispersion to $\sigma^2_{\text{int}}=0.1$ to be consistent with Pantheon \citep{Scolnic18}. We did not decompose intrinsic scatter into components incorporating dependencies on $\alpha$ and $\beta$ \citep{Marriner2011}. Given this project's focus on studying the host bias before and after standardization, no attempt was made to incorporate host properties into equation \ref{eq:standardization}. Magnitude uncertainty due to peculiar velocity was estimated using an empty-universe approximation with a peculiar velocity of $\sigma_v=300$ km s$^{-1}$ for each SN \citep[][, Eq.~A4]{Davis2011}:
\begin{equation}
    \sigma_{\mu} = \frac{\sigma_v}{c}
    \frac{5}{\ln(10)}\bigg[\frac{1 + z_{\rm CMB}}{z_{\rm CMB}(1+\frac{z_{\rm CMB}}{2})} \bigg].
\end{equation}

\begin{figure}
    \centering
    \includegraphics[width=\linewidth]{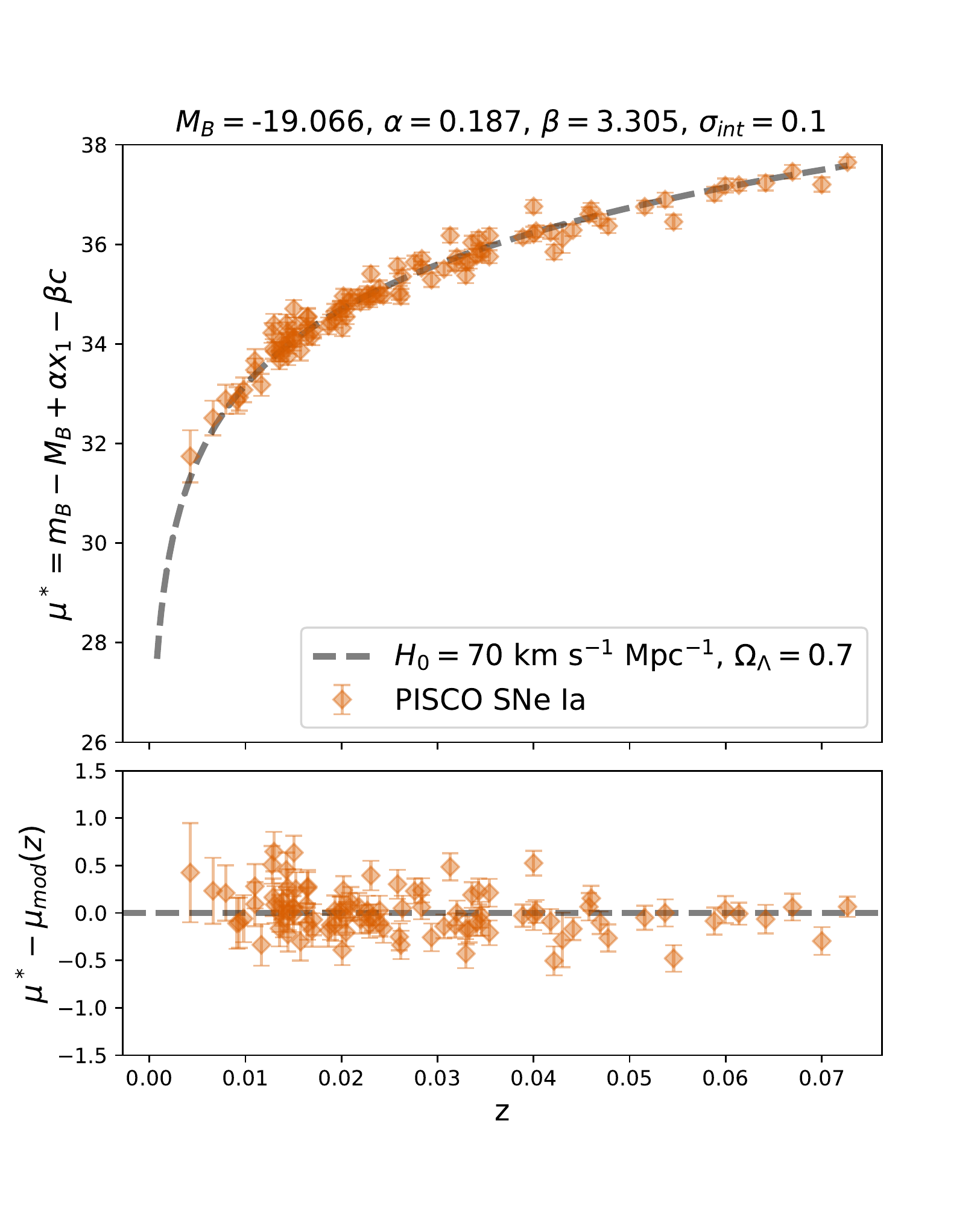}
    \caption{(top) Hubble diagram comparing SALT2-standardized PISCO SN~Ia distance moduli to a predicted flat $\Lambda$CDM cosmology with H$_0=70$ km s$^{-1}$ Mpc$^{-1}$ and $\Omega_{\Lambda}=0.7$ given as the dashed line.  Nuisance parameters $M_B$, $\alpha$, and $\beta$, and the fixed intrinsic SN Ia scatter $\sigma_{\text{int}}$ are given at the top of the top plot.  (bottom) Hubble residuals $\mu^* - \mu_{\text{mod}}(z)$ against redshift, with a dashed line at $\mu^* - \mu_{\text{mod}}(z)=0$ for reference.}
    \label{fig:hres_diagram}
\end{figure}

\subsection{Linear Regression and STAN}
\label{sec:methodology:regressions}
When comparing SFR estimates we used a Bayesian linear regression mixture model LinMix developed and described by \cite{KellyB2007} \footnote{\url{https://github.com/jmeyers314/linmix}}. This model takes into account uncertainties in both dependent and independent variables by modeling the true values of said variables as latent parameters described by a Gaussian mixture model.  Otherwise,  Ordinary Least Squares (OLS) linear regression implemented with SciPy\footnote{\url{https://www.scipy.org/}} was used where noted \citep{2020SciPy}. 

We used Stan\footnote{\url{https://mc-stan.org/}} to sample the posterior of a model between Hubble residuals and host property in place of a step function. 
The built-in Stan Hamiltonian Monte Carlo sampler approximates the Jacobian of model parameters, and since a step function features an undefined derivative at the step location, we instead used a shifted and scaled hyperbolic tangent (tanh) model:
\begin{equation}
    [\mu^* - \mu_\text{mod}(z)](x) = \Delta\mu \tanh{\frac{x - x_s}{\alpha}} + \Delta\mu_0
\end{equation}
where $x$ is the host property of interest (mass or sSFR). $\Delta\mu$ and $x_s$ parameterized the step size and step location, respectively with $x_s(M_{\odot}) \doteq M_s$ and $x_s(sSFR) \doteq sSFR_s$ for the mass and sSFR step location parameters. Parameter $\Delta\mu_0$ accounted for any systematic y-axis offset.  Parameter $\alpha$ determined the transition scale and was fixed to $\alpha=0.01$ to enforce a step-like trend, akin to the logistics model used by \cite{Scolnic18}, \cite{Brout2020}, and \cite{Popovic2021}.  Appendix \ref{app:tanh_model} details priors used and their respective motivations.

\section{Results and Analysis} \label{sec:analysis}

The robustness of stellar mass estimates is why a mass step is the most common characterization of the host-correlated bias in SN~Ia distance standardization \citep{Conley11, Betoule14, Scolnic18}.  But given that all host properties are is correlated with other host properties, it is not surprising that both SFR and  sSFR step-like biases have also been detected \citep{D_Andrea11, Rigault13, Rigault15, Rigault18}. In this work we use our available H$\alpha$ flux and UV photometry alongside optical spectra and photometry to estimated various mass and sSFR samples to determine if any chosen observation method or fitting technique leads to a significant change in the measured host bias.  

The analysis is presented with each subsection having both a summary of results followed by a discussion to make our analysis more digestible for the reader.

\subsection{Mass Estimate Comparisons}\label{sec:analysis:mass_comp_est}

\begin{figure}
    \plotone{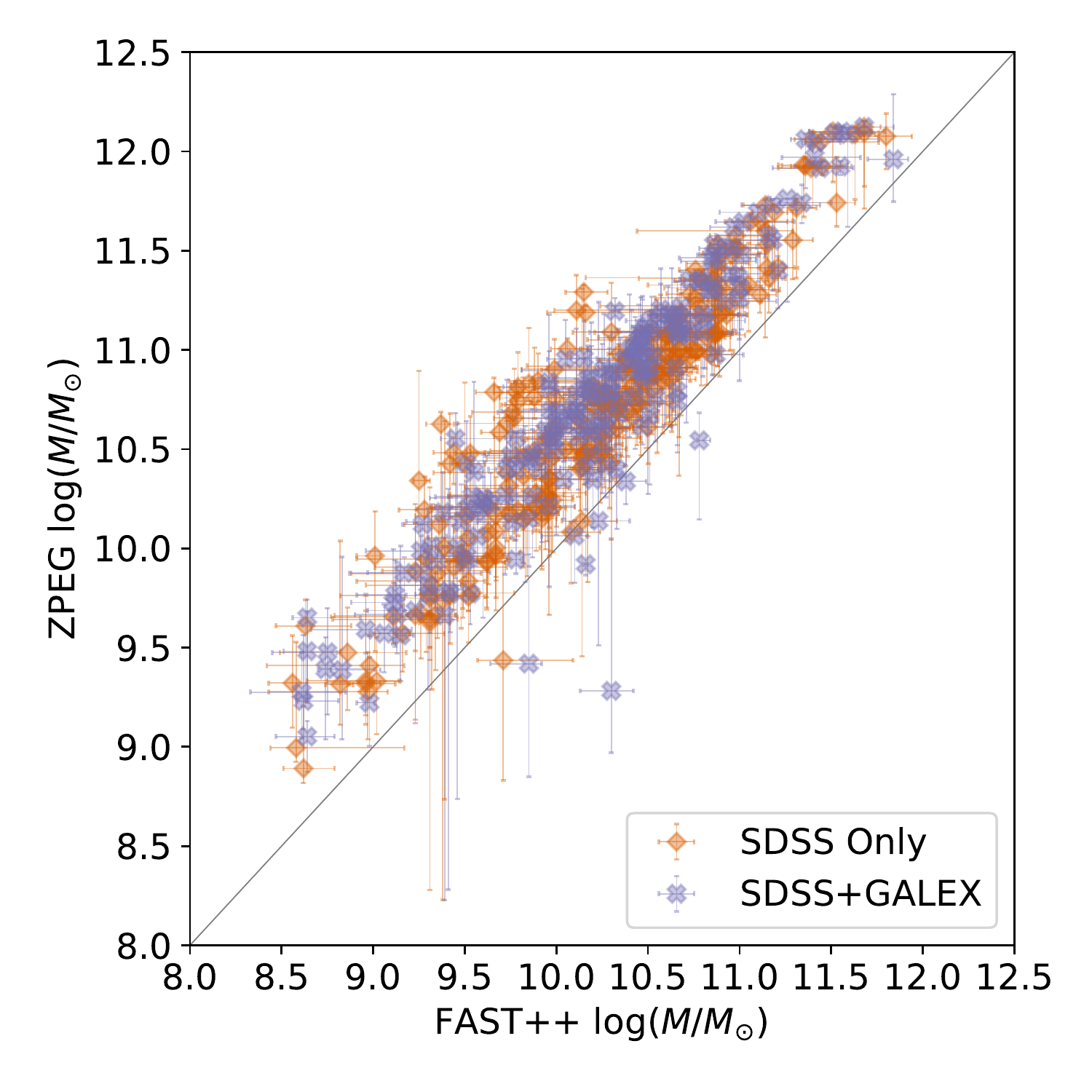}
    \caption{A comparison of stellar mass estimates between ZPEG and FAST++ for SDSS-only and SDSS+GALEX photometry. Error bars are taken directly from the respective fitting technique. A diagonal one-to-one grey line is included for reference.}
    \label{fig:mass_fppvszp}
\end{figure}

\begin{figure}
    \plotone{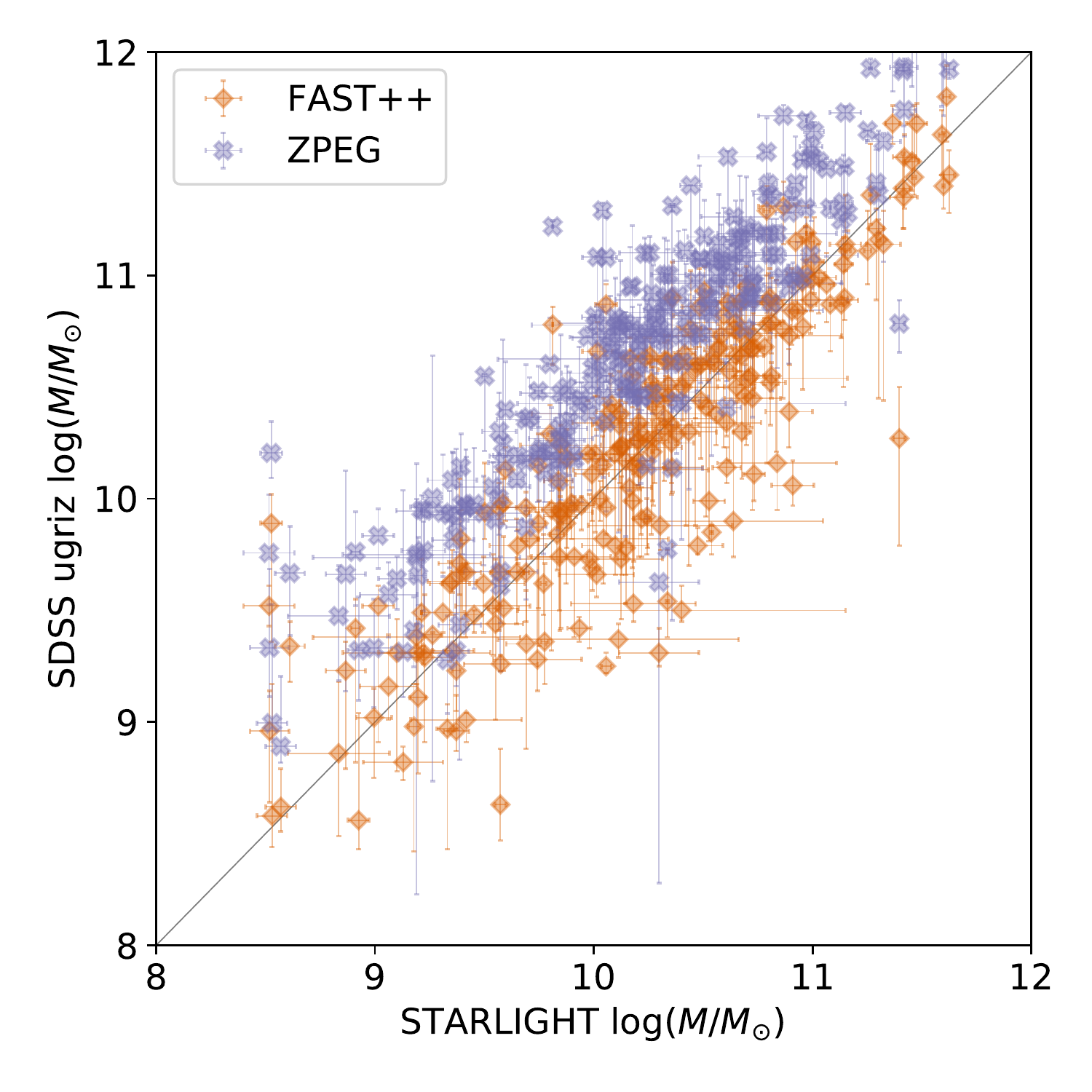}
    \caption{A comparison of mass estimate techniques. We found an offset of 0.47~dex between ZPEG and STARLIGHT mass estimates. Standard deviation between mass estimate differences were 0.28~dex for ZPEG (with the 16 erroneous estimates excluded) and 0.33~dex for FAST++.}
    \label{fig:starlight_mass_comp}
\end{figure}

Figure~\ref{fig:mass_fppvszp} compares fit mass estimates from FAST++ and ZPEG using SDSS and SDSS+GALEX photometry.  There was a median offset of  -0.43 dex (for SDSS only) and -0.54 dex (for SDSS+GALEX), largely consistent with that expected from differences between the Chabrier and Salpeter IMFs ($\sim$0.2 dex) and the stellar libraries BC03 and PEGASE.2 ($\sim$0.1 dex) \citep{Moustakas2013}.  The increase in median offset after including GALEX was likely driven by discrepancies between handling of young, massive stars in the respective libraries.

We compared mass estimates from FAST++ and ZPEG using SDSS optical \textit{ugriz} to STARLIGHT mass estimates using PISCO optical spectra.  Only 308 of 319 PISCO hosts had BC03+Chabrier IMF STARLIGHT mass estimates.  Of those 308 PISCO targets, very low redshift hosts Andromeda, M82, and NGC~6946 had failed mass estimates. Low redshift NGC~2276 was observed as three targets, each with failed mass estimates and were excluded. This reduced the usable STARLIGHT mass sample to 302 galaxies with STARLIGHT mass estimates.  Our requirement for mutual SDSS coverage reduced our working sample to 237 hosts.   In comparing mass estimates, we measured an average offset of 0.03 dex between the FAST++ and STARLIGHT mass samples with a standard deviation of 0.33 dex (Figure~\ref{fig:starlight_mass_comp}).  This small offset stemmed from the FAST++ SPS library and the SSP basis used by STARLIGHT both being constructed with a BC03 library and a Chabrier IMF.  A global median offset of 0.47~dex towards larger ZPEG mass estimates was caused by use of a Salpeter IMF and the PEGASE.2 spectral library for ZPEG \citep{Moustakas2013}. Mass offset standard deviation between the ZPEG and STARLIGHT mass samples was $0.28$~dex, slightly less than that measured for FAST++ mass values.

Mass estimates from optical photometry and IFS were mostly consistent with predicted global offsets. ZPEG mass estimates made with only SDSS photometry had systemically $\sim$0.15~dex higher values than their STARLIGHT or FAST++ counterparts --- this could be from the choice in library or dust model of ZPEG being biased towards older red stellar populations without UV constraints.  Random variation was consistent with SPS library uncertainties of $\sim$0.3~dex, with some further scatter arising from different aperture size and shape between PISCO IFS and our in-house photometry. 

\begin{figure}
    \centering
    \includegraphics[width=\linewidth]{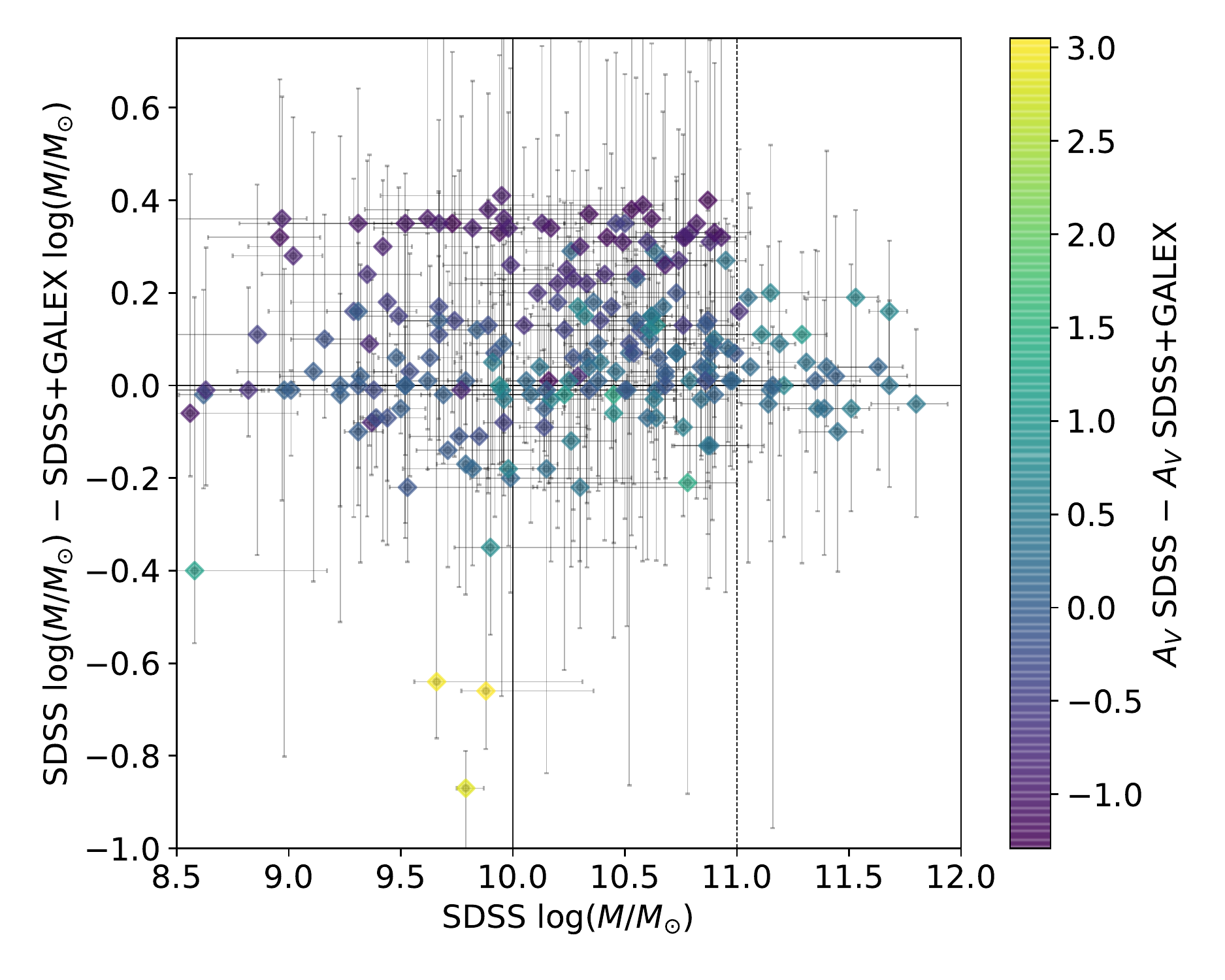}
    \caption{Absolute change in BC03 FAST++ stellar mass estimates after including GALEX photometry for 211 hosts. Three outliers exhibit mass increases of $>0.6$~dex, which would shift them over the fiducial mass step location marked with the vertical solid line.
    However we don't have light curves for the corresponding SNe~Ia and thus these hosts aren't actually used in the host galaxy bias analysis present here.  No mass shift magnitude greater than 0.2 dex was observed for massive hosts with $\log_{10}(M/M_{\odot}) > 11$, marked with the vertical dotted line. Note the correlation between differences in the best-fit $A_V$ and mass --- an example of the mass-age-dust degeneracy.}
    \label{fig:mass_sdssvssdss+galex}
\end{figure}

\begin{figure}
    \centering
    \includegraphics[width=\linewidth]{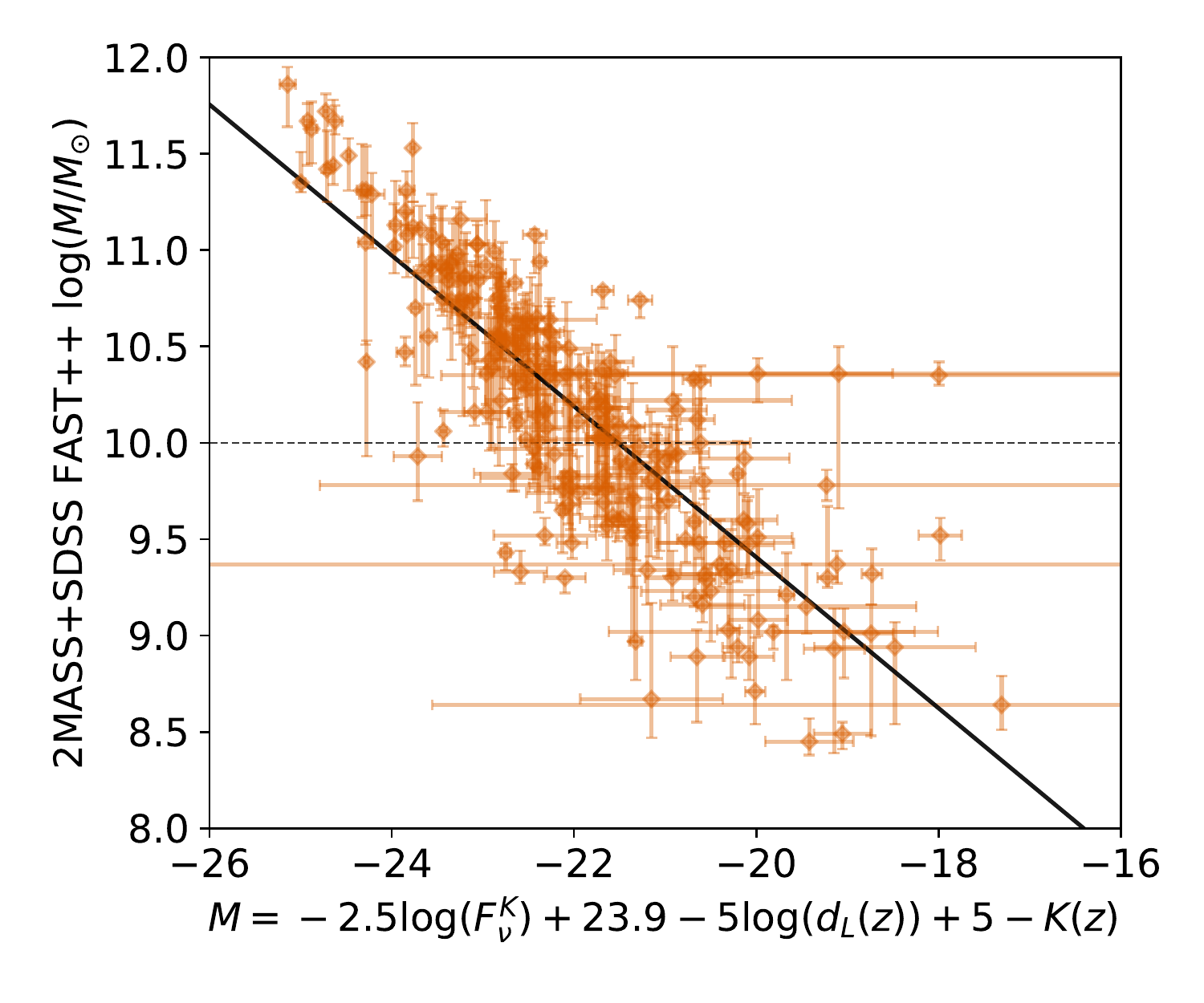}
    \caption{A comparison between \textit{K}--band absolute magnitude and FAST++ mass estimates determined using 2MASS+SDSS photometry for our PISCO sample.  The dashed black line references $\log_{10}(M/M_{\odot})=10$, the approximate location of the canonical mass step. The standard deviation of residual in mass estimates for galaxies above the dotted line is 0.35 dex compared to 0.46 dex for those below.  The solid black line is the linear OLS fit.}
    \label{fig:kmag_mass}
\end{figure}

\subsubsection{Effects of Incorporating UV Information}\label{sec:analysis:mass_comp_est:uv}

With FAST++ we compared mass estimates with and without GALEX NUV and FUV photometry. This required GALEX coverage and limited our sample to 211 galaxies from 237.  Adding UV information helped better constrain FAST++'s $A_V$ dust parameter, with average $A_V$ uncertainty decreasing by a factor of 2 from 0.6~mag to 0.3~mag. There was an appreciable anti-correlation between the change in fit $A_V$ and the change in fit mass (Pearson's r score of -0.74).  
Figure~\ref{fig:mass_sdssvssdss+galex} shows three clear outliers where mass increased by more than 0.6~dex once UV information was included: UGC~02134 and UGC~09165, both inclined spiral galaxies with clear UV signals, and 2MASXJ02305208, the smallest of a triple elliptical cluster. Note that both SN~Ia hosts UGC~09165 and 2MASXJ02305208 had mass estimates shift from above to below the canonical mass step location $\log_{10}(M/M_{\odot})=10$, but that these SNe~Ia lacked optical light curves and were not used in our mass step analysis (Section~\ref{sec:analysis:host_mass_bias_comp}).

 UV flux comes from active or recent star formation. An elliptical galaxy incorrectly described with a star forming model SED using only optical photometry can be correctly identified as elliptical once UV information is included, as was the case for 2MASXJ02305208. For UGC~02134 and UGC~09165, UV information provided further information about dust-obscured star formation, resulting in a lower stellar mass estimate after UV addition. 
Apart from these these outliers, remaining shifts in mass were within the 0.4~dex, with 52 estimates shifting with absolute value greater than 0.2~dex.  All but five of theses 52 hosts with absolute value shifted from higher to lower mass with UV information included.  This information helped break the color degeneracy for these 47 objects, with redness instead attributed to dust-obscured young stars with lower mass-to-light ratios as opposed to older red dwarf stars with high mass-to-light ratios, reducing mass estimates.

Nonetheless, aside from identifying galaxies that are forming no stars, mass estimation is largely insensitive to dust effects \citep{Conroy2013}.  This means that including UV information does not significant change mass estimates the vast majority of our sample (1.4\% of our 211 hosts for this subsample).  No hosts with SDSS mass $\log_{10}(M/M_{\odot}) > 11$ changed by more than 0.2~dex after including UV information.  Overall, we do  recommend including UV information when estimating mass given its ability to partially break the color degeneracy by distinguishing old stellar populations from dust-obscured young stars, but it does not significant change mass estimates for the sample. 

\subsubsection{Mass from NIR}

We also experimented with including 2MASS NIR information with SDSS photometry, finding mass estimates effectively unchanged. Average fit mass estimate uncertainty decreased for hosts with $\log_{10}(M/M_{\odot}) > 10$, though.  NIR photometry provided insufficient new information to constrain stellar mass beyond that already provided by optical photometry.

We considered the linear relationship between 2MASS \textit{K}-band magnitudes and FAST++ mass estimates using 2MASS and SDSS photometry, akin to the methodology used in \cite{Betoule14} (Figure~\ref{fig:kmag_mass}).
Two trends were present: 
(1) a linear relationship between absolute magnitude and mass estimate, with brighter absolute magnitudes being correlated with larger masses;
and
(2) an increase in the typical magnitude uncertainty for dimmer objects. The standard deviation of mass estimate residual was 0.35 dex below $\log_{10}(M/M_{\odot})=10$ and 0.46 dex above.
Neither relationship was surprising, but we explicitly mention these as the resulting trend is visibly and logically heteroscedastic --- dimmer objects will have larger observational errors, thus biasing lower-mass galaxies towards greater mass estimate uncertainties.
This should have no influence upon a step function model for a host bias correction, as the ordering of mass estimates would be effectively unchanged given the \textit{K}-mag trend with mass is clearly linear: a change in slope would have no impact upon said step model. 

\subsection{Host Mass Bias Comparison} \label{sec:analysis:host_mass_bias_comp}

\begin{figure}
    \centering
    \includegraphics[width=\linewidth]{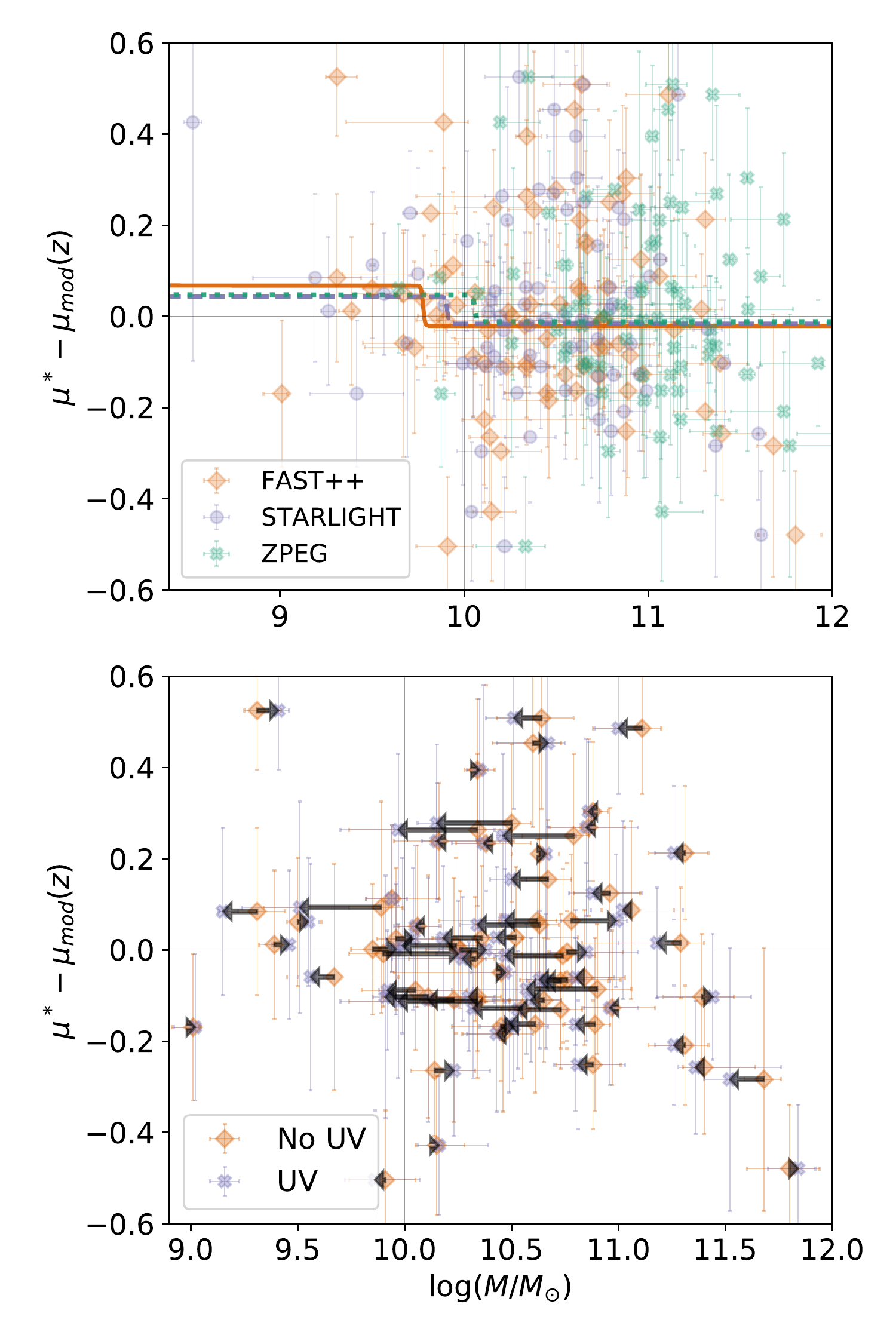}
    \caption{Comparison of 76 PISCO Hubble residuals to three sets of stellar mass estimates.  SDSS \textit{ugriz} photometry was used to calculate FAST++ and ZPEG mass estimates.  The bias towards high-mass SN~Ia hosts in the PISCO sample is apparent.  Tanh function fit results are given as a solid orange line for FAST++, a dashed green line for ZPEG, and a dotted purple line for STARLIGHT. The bottom panel is a comparison of 66 mass estimates from FAST++ excluding (orange diamonds) and including (purple x's) GALEX photometry from the base SDSS \textit{ugriz}. Despite noted shifts from high to low mass, both samples measured the same step function size, despite 10 fewer data points.  Dotted lines in both plots are for reference.}
    \label{fig:hres_mass_comp}
\end{figure}

\begin{deluxetable}{lrrr}
\tablecaption{Tanh model parameter results for mass samples.} \label{table:mass_step}
\tablehead{\colhead{Mass Sample} & \colhead{$M_s$} & \colhead{$\Delta \mu$} & \colhead{$\Delta \mu_0$}}
\startdata
76 FAST++ & $9.78\pm0.37$ & $-0.04\pm0.02$ & $0.02\pm0.02$\\
66 FAST++ & $9.73\pm0.39$ & $-0.04\pm0.03$ & $0.02\pm0.02$\\
66 FAST++ \& UV & $9.76\pm0.35$ & $-0.05\pm0.03$ & $0.02\pm0.02$\\
76 ZPEG & $10.06\pm0.53$ & $-0.03\pm0.02$ & $0.02\pm0.02$\\
76 STARLIGHT & $9.91\pm0.56$ & $-0.03\pm0.02$ & $0.01\pm0.02$\\
97 STARLIGHT & $9.83\pm0.33$ & $-0.04\pm0.02$ & $0.02\pm0.02$\\
\enddata
\end{deluxetable}

Hubble residuals for 76 SNe~Ia were compared against three mass estimates samples: FAST++ using SDSS, ZPEG using SDSS, and STARLIGHT using PISCO IFS.   These SN~Ia hosts had both SDSS coverage and available STARLIGHT mass estimates.  The decision to use only visible wavelength SDSS photometry for FAST++ mass estimates here were to maximize our usable sample and because PISCO SED only covers visible wavelengths.   Linear OLS regressors fit for all three mass samples found slope values consistent with a zero slope. Using our tanh step model implemented with Stan, we found $M_s=9.78 \pm 0.37$ and $\Delta \mu=-0.04 \pm 0.02$~mag for the FAST++ mass sample. For the STARLIGHT mass sample we found $M_s=9.91 \pm 0.56$ and $\Delta \mu=-0.03 \pm 0.02$~mag. With all available 97 STARLIGHT mass estimates the fit step size and location were more consistent with the 76 FAST++ mass sample at $M_s = 9.83 \pm 0.33$ and $\Delta \mu = -0.04 \pm 0.02$~mag.  
ZPEG fit parameters for the 76-host sample saw a less than 1$\sigma$-significant shift of 0.23~dex to $M_s = 10.06 \pm 0.53$ with a slightly reduced step size of $\Delta \mu=-0.03 \pm 0.02$~mag.  This shift toward a higher mass step location relative to the FAST++ mass sample was consistent with an expected $\sim$0.3~dex bias towards higher mass due to ZPEG's using the PEGASE.2 library and a Saltpeter IMF (Section~\ref{sec:analysis:mass_comp_est}). Also, the ZPEG and STARLIGHT step locations coincided with center of our mass step location bounds,  $[9, 11]$, albeit with ZPEG's step location being 0.15~dex higher than STARLIGHT's.  These parameter uncertainties were higher than their FAST++ counterpart by 0.16~dex and 0.19~dex for ZPEG and STARLIGHT, respectively, and were sufficient to span the allowed step location parameter space. See Table~\ref{table:mass_step} for a summary of $\tanh$ model results for our used mass subsamples. 

We found no evidence that using optical spectra versus optical photometry created any biased mass step. All step size variation between the 76-host mass samples were within 1$\sigma$ of each other.  Despite these two mass samples having different observation method and fitting technique, using all 97 STARLIGHT mass estimates produced fits results clearly consistent with the 76-host FAST++ mass sample fit parameters. We interpreted the increased step location uncertainty and reduced step size of the 76-host STARLIGHT sample fit as an artifact of our reduced sample size. The ZPEG mass step location was predicted to be greater than that fit for STARLIGHT and FAST++ mass samples (Section~\ref{sec:analysis:mass_comp_est}), but its reduced step size relative to FAST++ results and a less constrained step location could have affected the best-fit ZPEG step location.  A similar weakness in mass step signal for the 76-host ZPEG and STARLIGHT samples relative to the FAST++ sample was partly due to the lower scatter between ZPEG and STARLIGHT mass estimates as presented in Section~\ref{sec:analysis:mass_comp_est}.  Again, our reduced statistics were likely to blame for weak signal detection.

\subsubsection{Effects of Incorporating UV Information}
The 66 SN~Ia hosts with both SDSS and GALEX coverage had mass estimates compared both with and without UV photometry. As seen in Figure~\ref{fig:mass_sdssvssdss+galex}, below $\log_{10}(M/M_{\odot})<11$ there was a preferred shift to lower mass for a substantial portion of the PISCO sample once UV information was included.  This trend carried over for these 66 hosts, with arrows in the bottom plot of Figure~\ref{fig:hres_mass_comp} predominately pointing towards lower mass.  In particular, eight mass estimates changed by $2\sigma=0.28$ dex, with seven of the eight shifting to a lower mass once UV information was included. For the 66 mass estimates fit with GALEX information we found $M_s=9.73 \pm 0.39$ and $\Delta \mu = -0.04 \pm 0.03$~mag; mass estimates made only SDSS only fit for parameter values $M_s=9.76 \pm 0.35$ and $\Delta \mu = -0.05 \pm 0.3$~mag.  These were both relatively consistent and consistent with the previously discussed 76-host SDSS sample results.

None of the three PISCO hosts with mass changes greater than 0.6~dex were used in our mass step analysis, thus preventing an extreme shift in mass over the mass step location.  We observed an asymmetry in mass change direction after including UV information for many of our SN~Ia, but these mass changes were relatively small at $\sim0.4$~dex and did not translate to notable qualitative change in the step model parameter values or uncertainties. Adding UV information caused an insignificant 0.33-$\sigma$ change in mass step from -0.04~mag to -0.05~mag.  Although we cannot rule out the possibility of occasional SN~Ia host galaxies being misidentified when UV information is excluded, the majority of hosts had largely stable mass estimates with or without UV information (98.5\% in our 211 sample).  Regardless, we recommend mass estimates be made with UV information included where possible to prevent potentially large mass estimate errors.

\begin{figure*}
    \centering
    \includegraphics[width=7in]{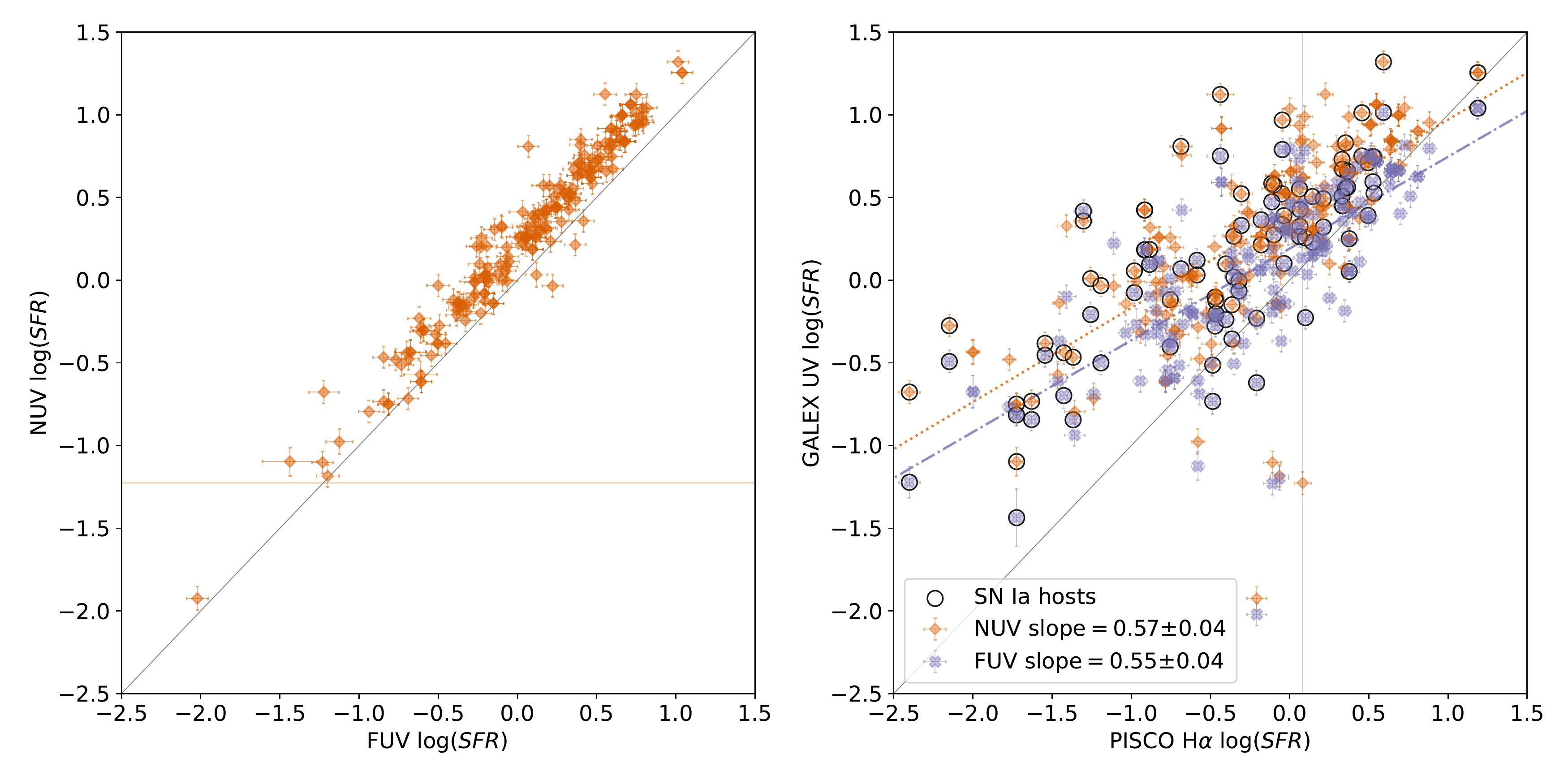}
    \caption{A comparison of global aperture SFR estimates using various techniques.  The left-hand plot compares FUV and NUV SFR estimates, which show very clear agreement with each other.  The right-hand plot compares FUV and NUV SFR estimates to H$\alpha$ SFR estimates.  The orange dotted line corresponds to NUV LinMix mean fit, while the purple dot-dashed line corresponds to FUV LinMix mean fit. Note that UV-calibrated SFR estimates include two sources of error: one from UV flux uncertainty and another 15\% fractional error from calibration uncertainty. Open circles correspond to the PISCO SN~Ia subsample.}
    \label{fig:sfr_global}
\end{figure*}

\subsection{SFR Comparison} \label{sec:analysis:sfr_comp}

 H$\alpha$ SFR estimates were calculated using a linear relationship provided by \cite{Calzetti2013}.  For consistency, a similar linear relationship for UV luminosity was also used to estimate UV SFR values. Using a single parameter star formation history for both ZPEG and FAST++ resulted in poorly constrained template SFR values.  As such, we excluded template SFR samples from the remainder of our analysis.

H$\alpha$ and UV SFR values were estimated using relationships from \cite{Leitherer1999} recalibrated for a Chabrier IMF.  H$\alpha$ SFRs were calibrated for a stellar mass range of 0.1--100~M$_{\odot}$ and star forming timescale $\tau \geq 6$~Myr.  Type-B recombination was assumed with $T_e=10^4$K and $n_e=100$ cm$^{-1}$:
\begin{equation}
    \text{SFR}({\text{H}\alpha}) = (0.88)~5.5 \times 10^{-42}L(\text{H}\alpha)
\end{equation}
where the factor of 0.88 accounts for our use of a Chabrier IMF instead of a Salpeter IMF and $L(\text{H}\alpha)$ is the luminosity calculated from the observed flux and luminosity distance.  \cite{Galbany2018} corrected H$\alpha$ flux for nebular host attenuation using measured versus theoretical H$\alpha$/H$\beta$ ratios. 

We used a UV SFR calibration valid for stellar mass range of 0.1-100~M$_{\odot}$ and star forming timescale $\tau = 10$~Myr:
\begin{equation}
    \text{SFR}({\text{UV}_{i}}) = (0.88)~4.3 \times 10^{-47}\lambda_i L(\text{UV}_{i})
\end{equation}
where $i$ indexes the NUV and FUV effective wavelengths.  For both equations, $L$ has units ergs s$^{-1}$.  STARLIGHT $A_V$ values were used to dust-correct UV flux via a Fitzpatrick dust extinction law with $R_V=3.1$, consistent with STARLIGHT's dust treatment described in Subsection~\ref{sec:methodology:methodology:gal_param_est:inversion_Fitting} \citep{fitzpatrick1999}. Note that UV-calibrated SFR estimates carried a further 15\% fractional error added in quadrature with luminosity uncertainty \citep{Calzetti2013}.

Four SN Ia hosts lacked H$\alpha$ flux measurements, excluding these from further analysis.  We also note that SN~Ib SN2003i's host IC2481 had erroneously large FUV flux uncertainty, as seen in both plots of Figure~\ref{fig:sfr_global}.

The left-hand plot from Figure~\ref{fig:sfr_global} demonstrated expected consistency between our NUV and FUV SFR estimates. There was a global median offset of 0.22 dex towards higher NUV SFR estimates, an artifact in the chosen SFR calibration's explicit linear dependence on wavelength that is partly accounted for by the mentioned 15\% fractional error added to our estimates.  Indeed, the offset remained unchanged when UV SFR estimates uncorrected for attenuation were instead used. 

\begin{figure}
    \centering
    \includegraphics[width=\linewidth]{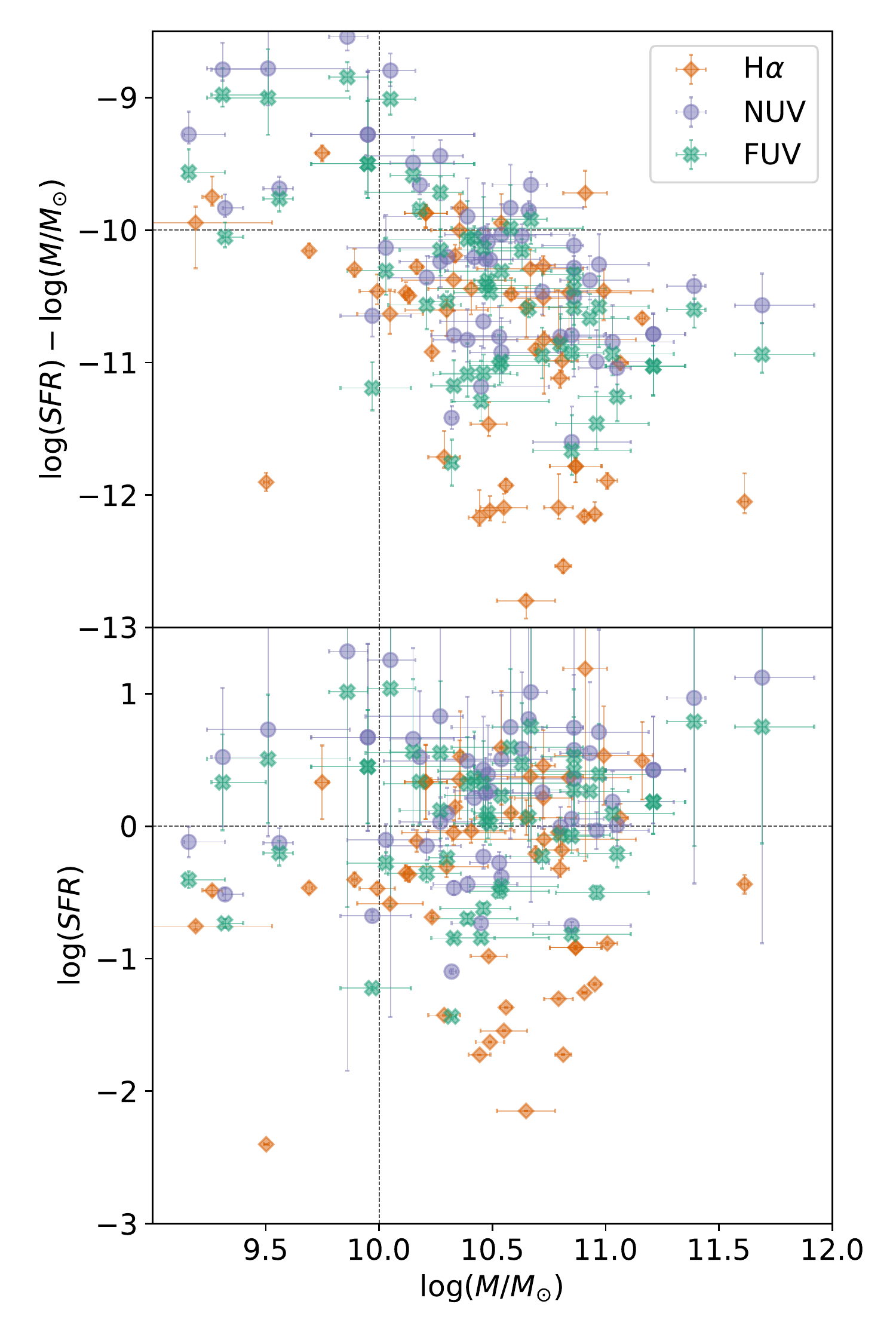}
    \caption{The top plot compares sSFR to stellar mass for all three flux-calibrated SFR samples for global aperture for our 51 host SN~Ia subsample.  The bottom plot similarly compares SFR to stellar mass. The lack of UV sSFR values below -12~dex is apparent. Dashed lines are for reference.}
    \label{fig:sfrssfr_mass}
\end{figure}

In the right-hand plot of Figure~\ref{fig:sfr_global}, we compared  UV-calibrated SFR estimates to H$\alpha$-calibrated SFR estimates from PISCO spectra.  To account for uncertainties along both axes we used LinMix to regress these UV samples against our H$\alpha$ sample.  Said regressions both had fit slopes less than one; OLS linear regression results were nearly identical to the mean LinMix model parameters. When UV photometry was uncorrected for dust, UV-H$\alpha$ SFR regression slopes were shallower at a 2.5$\sigma$ significance . An physical source of the observed scatter resulted from UV flux tracing SFR timescales of 10-100~Myr; nebular emissions such as H$\alpha$ trace a near-instantaneous ($<10$~Myr) SFR timescale \citep{Kennicutt1998}. We attempted to alleviate this by using a UV SFR relationship recalibrated for a shorter star formation timescale ($\approx 10$~Myr), but such recalibration translates to only a global offset in $\log_{10}(SFR)$ estimates. Readily apparent in Figure~\ref{fig:sfrssfr_mass} is said star formation timescale reducing the fit slope in SFR sample comparison, with UV SFR values being consistently larger than H$\alpha$ SFR estimates in the range $\log_{10}\text{SFR}(\text{H}\alpha)< -1$ $M_{\odot}yr^{-1}$.  This was  consistent with H$\alpha$ and UV flux both capturing very early star formation, but only UV flux capturing B-type stellar flux contribution after ionizing O-type stars have died out. 
Difference in our elliptical UV photometry apertures and PISCO's hexagonal apertures contributed to the observed scatter as well.

\subsection{sSFR Bias Comparison} \label{sec:analysis:ssfr_bias_comp}

\begin{deluxetable}{lrrr}
\tablecaption{Tanh model results for sSFR samples.} \label{table:ssfr_step}
\tablehead{\colhead{sSFR Sample} & \colhead{$sSFR_s$} & \colhead{$\Delta \mu$} & \colhead{$\Delta \mu_0$}}
\startdata
51 H$\alpha$ & $-10.96\pm0.48$ & $0.05\pm0.03$ & $0.03\pm0.02$\\
73 H$\alpha$ & $-10.83\pm0.73$ & $0.02\pm0.03$ & $0.02\pm0.02$\\
51 H$\alpha$ \& FAST++ & $-10.42\pm0.41$ & $0.06\pm0.03$ & $0.03\pm0.02$\\
51 FUV & $-10.28\pm0.52$ & $0.06\pm0.02$ & $0.04\pm0.02$\\
51 NUV & $-10.14\pm0.42$ & $0.06\pm0.02$ & $0.04\pm0.03$\\
\enddata
\end{deluxetable}

\begin{figure}
    \centering
    \includegraphics[width=\linewidth]{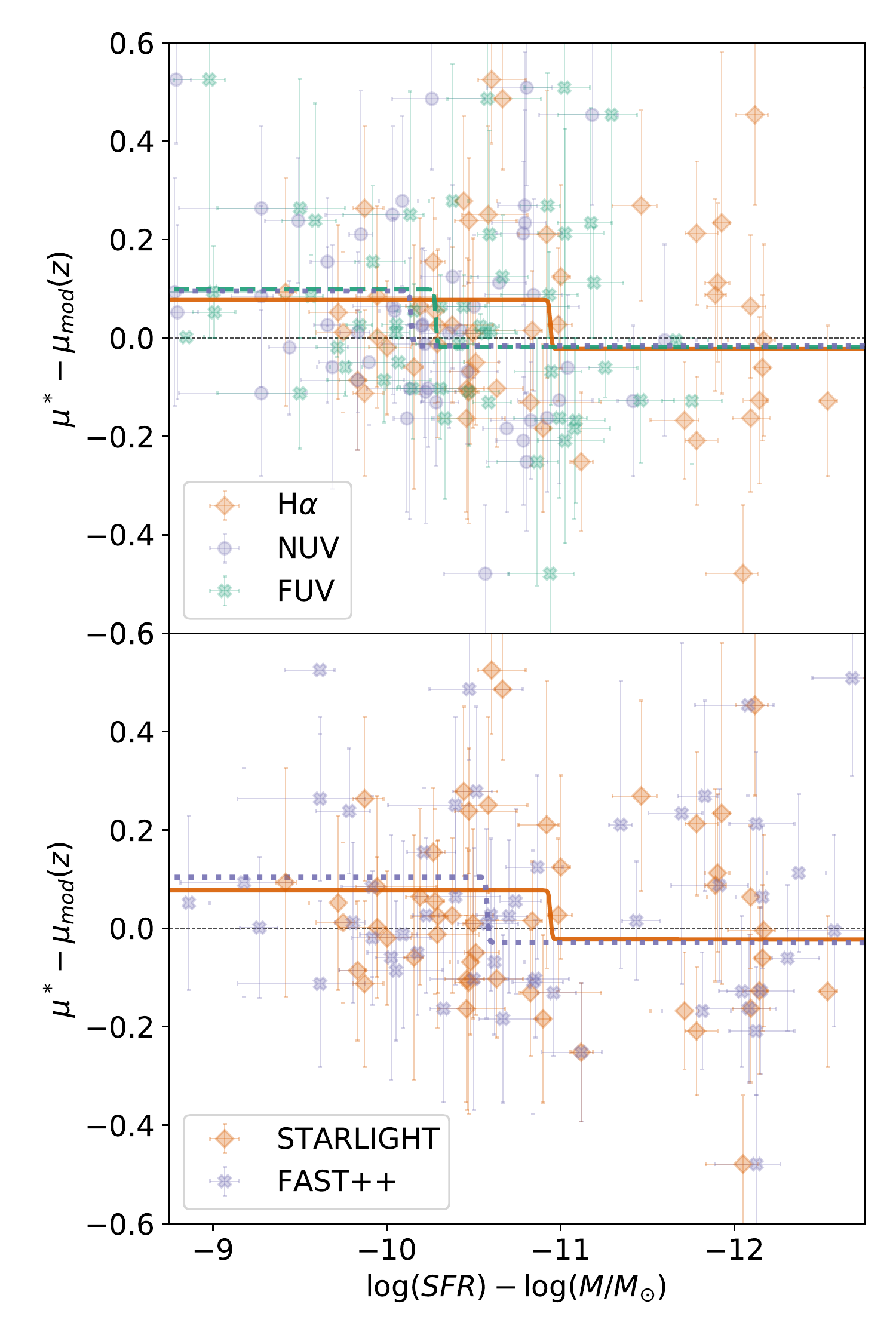}
    \caption{The top plot is a comparison of differing sSFR estimates to Hubble residuals for 51 SNe~Ia.  The bottom plot is a comparison the H$\alpha$ sSFR estimates using STARLIGHT or FAST++ mass estimates (consistent and inconsistent apertures, respectively).  sSFR tanh model fits are given in the top plot for H$\alpha$ (solid orange), NUV (dotted purple), and FUV (dashed green).  The bottom plot compares H$\alpha$ sSFR normalized using STARLIGHT (orange diamonds) and FAST++ (purple x's) masses. The solid orange line is the best fit tanh model for sSFR values from STARLIGHT mass estimates, the dotted purple line is the tanh model result instead using FAST++ mass values. }
    \label{fig:hres_globalssfr}
\end{figure}

In the top plot of Figure~\ref{fig:hres_globalssfr} we took 51 Hubble residuals of SNe~Ia with PISCO H$\alpha$ and GALEX FUV+NUV photometry coverage to three sSFR samples: FUV and NUV SFRs normalized by FAST++ mass estimates, and H$\alpha$ SFR normalized by STARLIGHT mass estimates.  FAST++ mass estimates were calculated with GALEX and SDSS photometry.  Linear OLS fits gave consistent slopes relative to each sSFR sample and were each statistically consistent with a zero slope.  With our Stan-implemented tanh model we found all three 51 host sSFR samples fit for slightly larger (and positive) step sizes than our mass step results with H$\alpha$ giving $\Delta \mu=0.05 \pm 0.02$, NUV giving $\Delta \mu=0.06 \pm 0.02$, and FUV giving $\Delta \mu=0.06 \pm 0.02$.  NUV and FUV step locations were $\text{sSFR}_s=-10.14 \pm 0.42$ and $\text{sSFR}_s =-10.28 \pm 0.52$, respectively. The fit H$\alpha$ step location was at a 0.5~dex lower sSFR with $\text{sSFR}_s =-10.96 \pm 0.41$. Normalizing H$\alpha$ sSFR values with FAST++ mass estimates gave a step location closer to its UV sSFR counterparts with $\text{sSFR}_s =-10.48 \pm 0.52$ and $\Delta \mu=0.06 \pm 0.03$~mag. Using instead all available 73 SN~Ia hosts with global H$\alpha$ flux to fit H$\alpha$ sSFR values against Hubble residuals saw the step size inexplicably drop a 1$\sigma$-significant shift by 0.03~mag to only $\Delta \mu=0.02 \pm 0.02$, with the step location parameter fitting for $\text{sSFR}_s =-10.83 \pm 0.73$. This step location was effectively unconstrained within the bounds enforced on the sampler. See Table~\ref{table:ssfr_step} for a summary of all sSFR tanh model results.

\begin{figure}
    \centering
    \includegraphics[width=\linewidth]{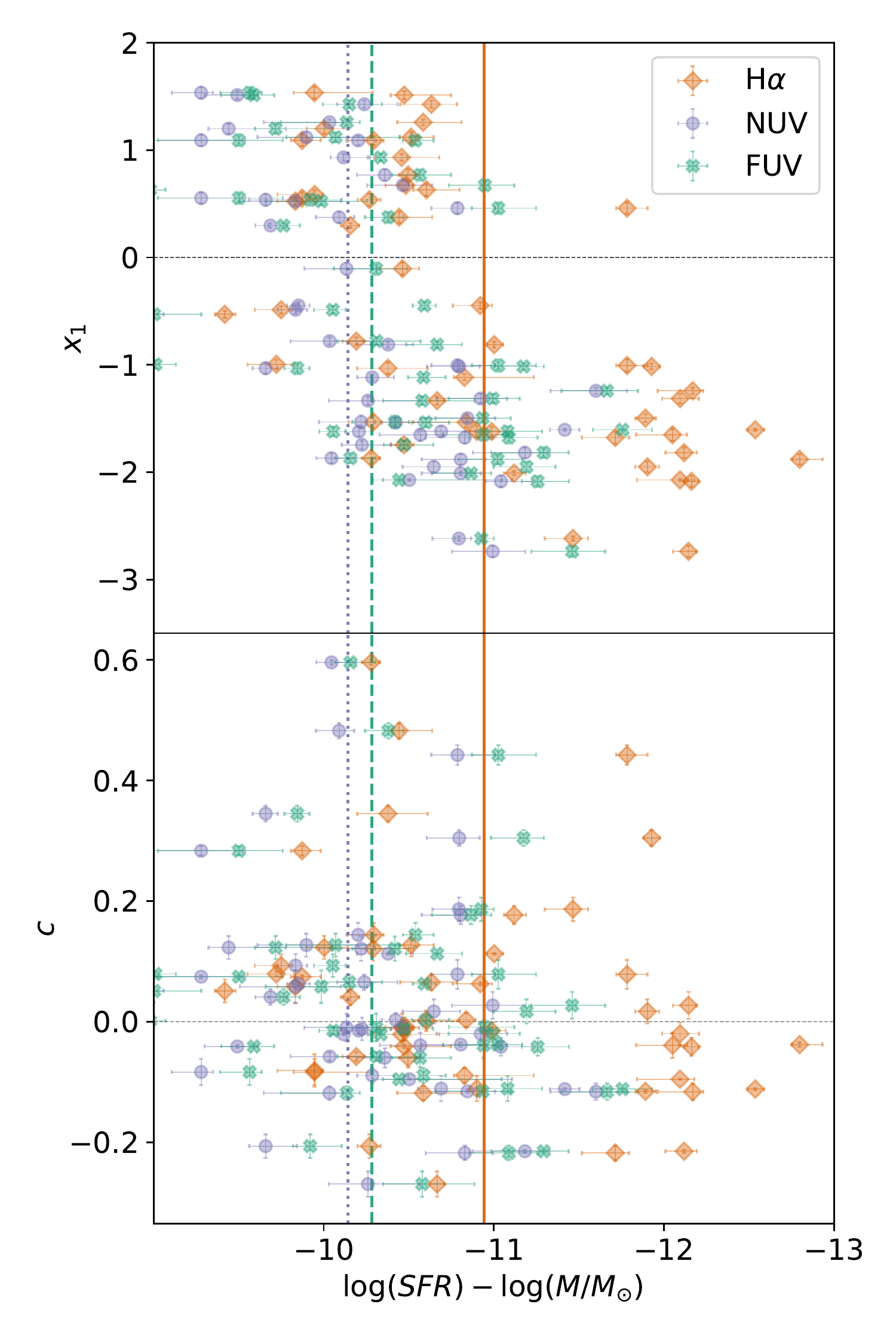}
    \caption{A comparison of global sSFR estimates to SALT2 parameters $x_1$ (top) and $c$ (bottom) for 51 SNe~Ia. Best-fit Hubble residual step locations are given by the vertical lines: H$\alpha$ corresponds to solid orange, NUV corresponds to dotted purple, and FUV corresponds to dashed green.  FUV and NUV step locations overlap. The discrepancy in lower sSFR values between H$\alpha$ and UV sSFR values is clear in both plots.}
    \label{fig:x1_c_globalssfr}
\end{figure}

In Figure~\ref{fig:x1_c_globalssfr} we compared SALT2 parameters $x_1$ and $c$ to the same three sSFR samples, marking with vertical lines the best-fit sSFR step location for each of the three samples.  Obvious in this plot were the lowest H$\alpha$ sSFR estimates being less than the lowest UV sSFR estimates by an order of magnitude. To see if the SN~Ia host bias was dependent on SFR tracer, we calculated the rank correlations coefficient Spearman's $\rho$ of the three sSFR samples against SALT2 $x_1$.  A relative comparison of these $\rho$ values demonstrated uniformity with modest rank correlation with $\rho=0.61$ for H$\alpha$, $\rho=0.66$ for NUV, and $\rho=0.62$ for FUV sSFR samples, providing tempered support for star formation epoch (10~Myr versus ~100~Myr) being an insignificant factor in our host bias measurement.   Interestingly, when repeated with UV SFR estimates uncorrected for dust, both NUV and FUV Spearman's $\rho$ values decreased to $\rho=0.55$ and $\rho=0.54$, highlighting the important of dust corrections when measuring the SN~Ia host bias. 

H$\alpha$~sSFR step size parameters had uncertainties within 2$\sigma$ significance of a zero step fit, but  FUV and NUV sSFR step size parameters fit with a 3$\sigma$ significance from a zero step.  Despite a modest sample size of 51 hosts, this 3$\sigma$ significance was the the best of any step models we fit in our analysis.  The combination of SFR and stellar mass provided a stronger and larger step signal than mass alone, consistent with both \cite{Rose2020} and \cite{Rigault18}.
The mass sample chosen to normalize H$\alpha$ SFR values influenced sSFR step location, with H$\alpha$ SFR normalized by FAST++ mass values having a step location closer to UV sSFR estimates by 0.05~dex. This observed sSFR step location shift was an order of magnitude larger than the corresponding 0.05~dex median mass offset between these 51 FAST++ and STARLIGHT mass values. The lower minimum H$\alpha$ SFR (as seen in Figures \ref{fig:sfrssfr_mass} and \ref{fig:x1_c_globalssfr}) explained the remaining difference between the the H$\alpha$ and UV sSFR step locations.  The observed step location change for NUV and FUV sSFR samples was a consequence of an explicit wavelength dependence in our SFR UV calibration.
There was only a slight increase in step size from H$\alpha$ to UV sSFR samples.  This less than 1$\sigma$ change was statistically insignificant and provided no evidence that tracing differing star formation timescales influenced sSFR step size.
Using all available 73 H$\alpha$ sSFR values normalized by STARLIGHT did result in weak step size detection well within 1$\sigma$ of a zero step. When fitting for the mass step using only these 73 PISCO hosts, we found an unconstrained mass step location, evidence that this particular PISCO host subset simply had a weak to nonexistent mass step signal. 

We found no evidence that observation method nor fitting technique influenced sSFR step size.
Lending support to this conclusion, the host bias strength inferred from Spearman's $\rho$ rank correlations between $x_1$ and sSFR samples were very similar.
The sSFR step location was partly influenced by SFR tracer choice.
Different SFR tracers measure different star formation ages
H$\alpha$-based SFR tracks young O-type stars, while UV-based SFR includes a wider range massive stars.

\section{Conclusion} \label{sec:conclusion}
This project sought to determine whether different observation methods or fitting techniques create or change the observed SN~Ia host bias in the SN~Ia standardized magnitudes.
When we examined subsets with more complete complementary photometry, the reduced statistics resulted in all mass step sizes being at or within a 2$\sigma$ significance with respect to a zero step, ranging from -0.03$\pm$0.02~mag to -0.04$\pm$0.02~mag. 
Mass step size values for our three optical wavelength mass samples (FAST++ and ZPEG with SDSS, STARLIGHT with PISCO) were all relatively consistent, being within a 1$\sigma$ significance of each other (Table~\ref{table:mass_step}).
The mass step location, but not size, varied under different IMF and stellar spectra choice.
Including GALEX UV when fitting for photometric mass estimates with FAST++ led to an insignificant 0.33$\sigma$ change in mass step size and no discernible change in corresponding step location.  Not including UV information could ostensibly influence mass step measurement, but that was not the case for our used PISCO subsample.  Given that this could matter, including available UV information during SPS fitting is the best practice.

The dust-corrected sSFR samples were fit against Hubble residuals for a 51-host subsample with both UV and SDSS coverage alongside available H$\alpha$ flux measurements (Table~\ref{table:ssfr_step}). Physically different star formation timescales traced by H$\alpha$ and UV flux sourced a H$\alpha$ sSFR step location $\sim$0.7~dex lower than UV sSFR step locations. Step size parameter values for these sSFR samples were all within 1$\sigma$ of each other. 
The sSFR steps from both FUV or NUV sSFR samples were the most statistically significant of all our step model runs at 3$\sigma$, with a step size of 0.06$\pm$0.02~mag.
Alternatively, using all available 73 PISCO SN~Ia hosts with H$\alpha$ flux produced the smallest step size of any model fit, being more than a factor of two less than the corresponding 51 host sample H$\alpha$ step size parameter value and within a 1$\sigma$ significance of a zero step.
With the 51 PISCO sample across all three sSFR samples the sSFR step sizes were clearly consistent for all observation methods and fitting techniques.  Thus, we concluded that the methodology or technique choice had no significant effect on sSFR step size measurements.  
Indeed, for this project only a particular change in sample size led to a discernible change in sSFR step parameter value for our particular data set.  

We undertook this study expecting to find that these differences would matter for the current state of SN~Ia cosmology.  We found that they did not in the presently available sample, but we nevertheless urge continued attention by the SN cosmology community to advances in our understanding of galaxy properties.
Given the complex nature of the SN~Ia host bias problem, it is paramount that galaxy community resources be researched and utilized to reach a resolution in the future and to be mindful of coming developments in both SED fitting and SPS research.
\acknowledgments
This work was supported in part by the US Department of Energy Office of Science under DE-SC0007914.
The authors thank the referee for their time, patience, and valued input in reviewing this paper in its submissions.  
The authors thank the Carnegie Supernova Project II for generously sharing the SALT-2 fits for unpublished SNe~Ia whose hosts were part of the PISCO sample.  The authors would also like to thank Dr. Alex Kim for both the introduction to and guidance using Stan.
L.G. acknowledges financial support from the Spanish Ministry of Science, Innovation and Universities (MICIU) under the 2019 Ram\'on y Cajal program RYC2019-027683 and from the Spanish MICIU project PID2020-115253GA-I00.

Based on observations collected at the Centro Astronómico Hispano-Alemán (CAHA) at Calar Alto, operated jointly by Junta de Andalucía and Consejo Superior de Investigaciones Científicas (IAA-CSIC).
Funding for SDSS-III has been provided by the Alfred P. Sloan Foundation, the Participating Institutions, the National Science Foundation, and the U.S. Department of Energy Office of Science. The SDSS-III web site is http://www.sdss3.org/.

SDSS-III is managed by the Astrophysical Research Consortium for the Participating Institutions of the SDSS-III Collaboration including the University of Arizona, the Brazilian Participation Group, Brookhaven National Laboratory, Carnegie Mellon University, University of Florida, the French Participation Group, the German Participation Group, Harvard University, the Instituto de Astrofisica de Canarias, the Michigan State/Notre Dame/JINA Participation Group, Johns Hopkins University, Lawrence Berkeley National Laboratory, Max Planck Institute for Astrophysics, Max Planck Institute for Extraterrestrial Physics, New Mexico State University, New York University, Ohio State University, Pennsylvania State University, University of Portsmouth, Princeton University, the Spanish Participation Group, University of Tokyo, University of Utah, Vanderbilt University, University of Virginia, University of Washington, and Yale University.

This publication makes use of data products from the Two Micron All Sky Survey, which is a joint project of the University of Massachusetts and the Infrared Processing and Analysis Center/California Institute of Technology, funded by the National Aeronautics and Space Administration and the National Science Foundation.

Both pandas and NumPy were constantly utilized throughout this project \citep{numpy2020, 2020pandas}.

\software{
    FAST++\footnote{\url{https://github.com/cschreib/fastpp}},
    STARLIGHT\footnote{\url{http://www.starlight.ufsc.br}},
    ZPEG\footnote{\url{http://imacdlb.iap.fr/cgi-bin/zpeg/zpeg.pl}},
    Python\footnote{\url{http://python.org}}, 
    AstroPy~\citep{astropy13}\footnote{\url{http://www.astropy.org}},
    NumPy\footnote{\url{http://www.numpy.org}}, 
    SciPy\footnote{\url{http://www.scipy.org}},
    STAN{\footnote{\url{https://mc-stan.org/}}},
    pandas{\footnote{\url{https://pandas.pydata.org/}}}
}
\clearpage
\appendix
\section{Issues with 1~kpc Photometry with GALEX}
\begin{figure}
    \centering
    \includegraphics[width=\linewidth]{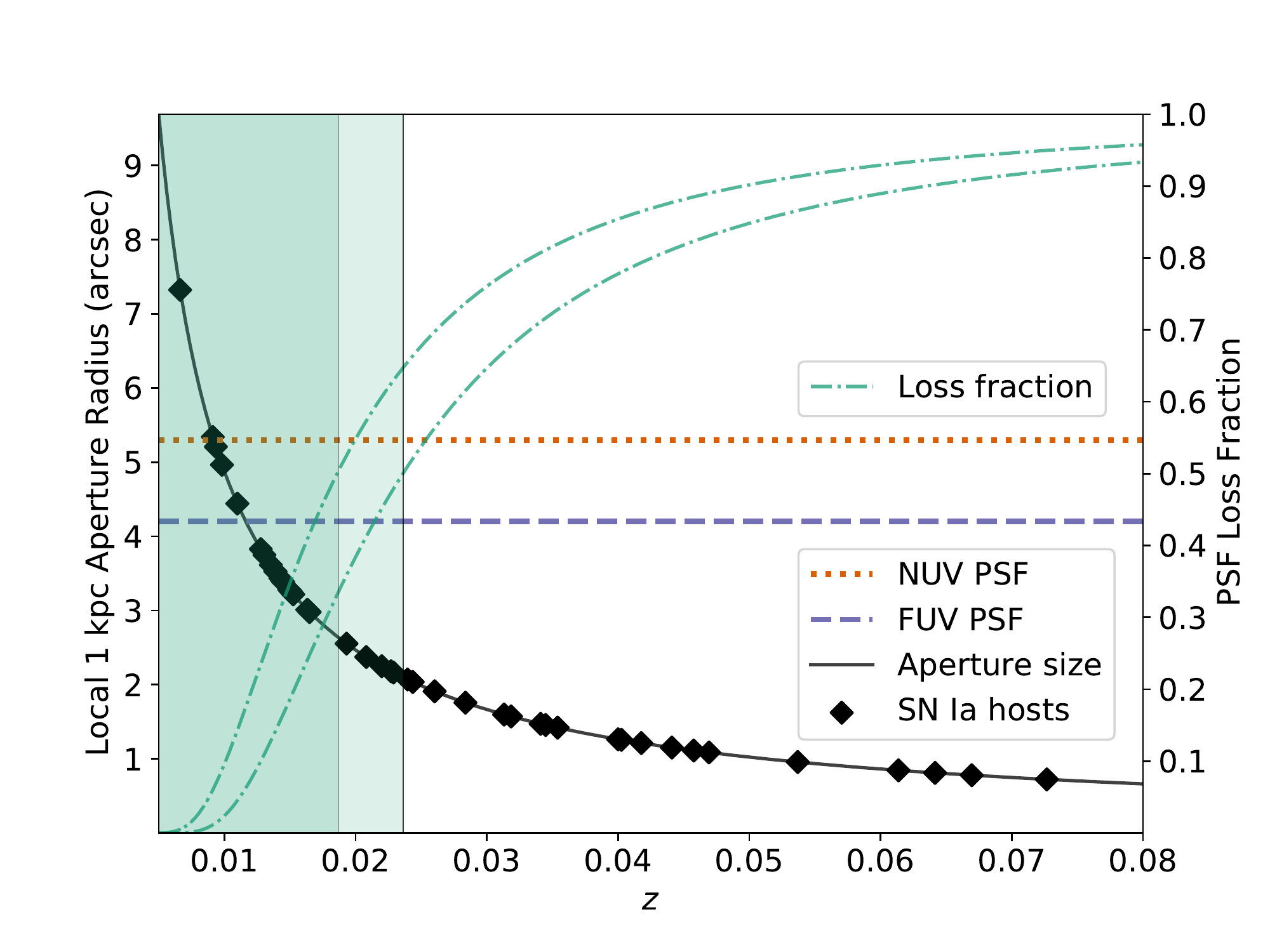}
    \caption{The solid monotonically-decreasing curve is the angular size of 1~kpc circular apertures in arc-seconds as a function of redshift $z$.  Black diamonds correspond to the PISCO SNe~Ia with `usable' 1~kpc GALEX photometry.  The horizontal orange dotted line is GALEX NUV average PSF size of 5.3\arcsec, while the horizontal purple dashed line is GALEX FUV average PSF of 4.2\arcsec.  The monotonically-ascending teal dot-dashed curves approximates a central point source's NUV or FUV flux loss fraction as the 1~kpc aperture size decreases with redshift. Teal shaded regions cover apertures with less than 50\% NUV (light+dark) and FUV (dark-only) point source flux loss.}
    \label{fig:galex_psf}
\end{figure}

We initially planned to perform 1~kpc circular aperture photometry around our SNe~Ia to compare local UV and H$\alpha$ SFR estimates, but the large GALEX PSF proved cumbersome to work with.
For pulse height bin values above 10 the PSF FWHM averaged at 5.3" or the NUV band and 4.2" for the FUV band \citep{Morrissey2007}. The PSF for GALEX NUV and FUV images was notably asymmetric for many observations. Thus, we excluded local 1~kpc aperture photometry from this project as these apertures' angular sizes were almost always less than average PSF as shown in Figure~\ref{fig:galex_psf}.

To get an idea of flux loss, we assumed a Gaussian profile for the PSF of a centralized point source and calculated the fractional flux loss as 1~kpc apertures decreased in size with increasing redshift.  The left and right teal curves in Figure~\ref{fig:galex_psf} correspond to point source NUV and FUV flux loss fraction with redshift, respectively.  Such calculations were crude approximations of extended structure, but the rapid blurring of UV flux with increasing redshift was dramatically evident ---~only targets with $z<0.025$ did not simultaneously suffer from $\sim50$\% point source flux loss and $\sim50$\% flux contribution from outside the 1~kpc aperture.  By $z=0.05$, you are effectively capturing none of the true flux originating from the 1~kpc aperture, instead almost entirely capturing photons originating from outside your aperture.  This approximation for a point source at the center of the aperture ignored the far more complex flux blurring for a resolved host.

Motivating our use of 1~kpc aperture photometry was to better approximate local properties such as SFR and stellar mass.  This was partly in response to GALEX's use in past SN~Ia host galaxy studies using apertures with radii as large as 4~kpc  \citep{Rigault15,Jones2015,Roman2018}. To treat UV flux measured from such an aperture as a proxy for local UV SFR is not physical.  Indeed, even a 1~kpc radius aperture is nearly four times the radius of the largest molecular clouds.  With these issues, it was obvious that GALEX could not be used to reasonably estimate SFR rates smaller than the entire host galaxy for all but the closest targets with $z<0.01$.

\section{Tanh Model} \label{app:tanh_model}
Stan's HMC sampler calculates a numerical Jacobian matrix of model parameters.  A true step function such as a Heaviside function has an undefined derivative at the specified step location which prevented model convergence for nearly all data sets we used.  Some works form the literature instead used a logistics function with a rapidly transitioning curve \citep{Scolnic18, Brout2020, Popovic2021}.  We chose to use a hyperbolic tangent function in place of a logistics function since $\tanh(0)=0$ and its antisymmetry, although such properties were lost when adding the y-axis offset parameter $\Delta \mu_0$. You can derive a a shifted tanh from a logistics function:
\begin{equation}
    \frac{1}{1 + e^{-2x}}=\frac{1}{2} + \frac{1}{2}\tanh{x}
\end{equation}

A relatively loose prior was placed on $\Delta\mu$ to prevent the sampler from getting lost at extreme $\Delta\mu$ values:
\begin{equation}
    \Delta\mu \sim \mathcal{N}(0, 0.5)
\end{equation}
The measured host bias step size in the literature is on the order of 0.1~mag or less.  Physically-motivated bounds were set to limit the sign of the step based on the host galaxy property being used: $\Delta\mu \in (-\infty, 0]$ and $\Delta\mu \in [0, \infty)$ for mass and sSFR, respectively. The direction of the host bias post-standardization is established, hence bounding the step direction. 

For both  The mass and sSFR step size parameters $x_s(M_{\odot}) \doteq M_s$ and $x_s(sSFR) \doteq sSFR_s$ both had a modest prior to deter the sampler from becoming stuck at parameter boundaries:
\begin{equation}
    x_s \sim \mathcal{N}(c, 1)
\end{equation}
where $c=10$ for $M_s$ and $c=-11$ for $sSFR_s$. Values for $c$ were the midpoints for the boundaries $M_s \in [9, 11]$ and  $sSFR_s \in [-12.5, -9.5]$. 
These bounds were placed to keep the fit step location within the actual mass or sSFR distribution for our subsamples used.

$\Delta \mu_0$ partly captured the systematic offset from zero for Hubble residual systematic offsets. As such, we used the informative prior:
\begin{equation}
    \Delta \mu_0 \sim \mathcal{N}\big(\langle \mu^* - \mu(z) \rangle,\text{SE}[\mu^* - \mu(z)]\big)
\end{equation}
where $\text{SE}[\mu^* - \mu(z)]$ is the standard error of the subsample Hubble residual  mean $\langle \mu^* - \mu(z) \rangle$. For our 76 host SN~Ia mass subsample $\langle \mu^* - \mu(z)u \rangle=0.010$~mag with $\text{SE}[\mu^* - \mu(z)]=0.003$~mag and for our 66 host mass subsample $\langle \mu^* - \mu(z) \rangle=0.012$~mag with $\text{SE}[\mu^* - \mu(z)]=0.003$~mag.  For our 51 host sSFR subsample $\langle \mu^* - \mu(z) \rangle=0.042$~mag with $\text{SE}[\mu^* - \mu(z)]=0.004$~mag and for our 73 host sSFR subsample $\langle \mu^* - \mu(z) \rangle=0.031$~mag with $\text{SE}[\mu^* - \mu(z)]=0.003$~mag.

Fixing the scaling term $\alpha$ and fitting our model with values increasing from $\alpha=0.01$ up to $\alpha=0.5$ had no appreciable effect on the sampled posterior.  If instead promoted to a free parameter and provided a uniform prior, $\alpha$ trended to large values, resulting in a pseudo-linear regression. Providing an informative exponential prior with $\lambda=0.3$ to keep $\alpha$ from trending past $\alpha = 1$ resulted in a near-identical posterior to those resulting from simpler models with fixed $\alpha$. Promoting $\lambda$ to a free hyperparameter with a inverse-gamma hyperprior again resulted in similar results, with the data only marginally influencing the $\lambda$ marginal posterior.  We opted to leave the scaling term fixed to $\alpha=0.01$ to provide a smooth step function approximation.

\begin{deluxetable}{lrrrr}
\tablecaption{Global mass in $\log_{10}(M/M_{\odot})$. FAST++, ZPEG masses limited to SDSS or SDSS+GALEX coverage. The first 25 entries are shown here. A full version of this table is available online.} \rotate \label{table:sfr}
\tablehead{\colhead{Galaxy} & \colhead{STARLIGHT Mass} & \colhead{Fast++ Mass} & \colhead{ZPEG Mass} & \colhead{Fast++ GALEX Mass}}
\startdata
2MASSJ00164451 & 10.094 & 10.200 & 10.765 & -- \\
2MASSJ00235669 & 9.653 & 9.790 & 10.192 & 9.770 \\
2MASSJ23243021 & 10.219 & 9.910 & 10.452 & 9.890 \\
2MASXJ22532475 & 9.891 & 9.960 & 10.348 & 10.300 \\
2MASXJ00234829 & 10.539 & 9.850 & 10.829 & 9.860 \\
2MASXJ01144386 & 10.434 & 10.300 & 10.723 & 10.430 \\
2MASXJ01403375 & 10.640 & 10.760 & 11.262 & 11.000 \\
2MASXJ02305208 & 10.013 & 9.660 & 10.787 & 10.300 \\
2MASXJ04424248 & 11.023 & -- & -- & -- \\
2MASXJ07192718 & 9.939 & -- & -- & -- \\
2MASXJ08374557 & 9.387 & 9.710 & 9.436 & 9.980 \\
2MASXJ09591230 & 9.826 & -- & -- & -- \\
2MASXJ10525434 & 10.147 & 9.780 & 10.787 & -- \\
2MASXJ12095669 & 10.285 & 10.410 & 10.816 & -- \\
2MASXJ12385810 & 10.638 & 9.900 & 10.844 & 9.840 \\
2MASXJ15024995 & 9.311 & 9.490 & 9.938 & -- \\
2MASXJ15393305 & 10.996 & 11.150 & 11.530 & 11.090 \\
2MASXJ15570268 & 9.839 & 10.080 & 10.082 & 10.080 \\
2MASXJ16065563 & 10.163 & 10.050 & 10.500 & 10.240 \\
2MASXJ16152860 & 9.807 & 9.950 & 10.207 & 10.030 \\
2MASXJ16301506 & 10.725 & 10.450 & 10.910 & 10.390 \\
2MASXJ17100856 & 10.202 & 10.170 & 10.605 & 10.160 \\
2MASXJ18242915 & 9.475 & -- & -- & -- \\
2MASXJ21352164 & 11.302 & 11.160 & 11.360 & 11.200 \\
2MASXJ23024668 & 10.675 & 10.640 & 11.161 & 10.650
\enddata
\end{deluxetable}

\begin{deluxetable}{lrrrrr}
\tablecaption{Global SFR in $M_{\odot}\text{yr}^{-1}$.\rotate \label{table:sfr}}
\tablehead{\colhead{Galaxy} & \colhead{FAST++ SFR} & \colhead{H$\alpha$ SFR} & \colhead{NUV SFR} & \colhead{FUV SFR} & \colhead{ZPEG SFR}}
\startdata
2MASXJ22532475 & 0.692 & 0.395 & 1.253 & 0.577 & 0.000 \\
2MASXJ00234829 & 141.254 & 3.910 & 20.854 & 10.341 & 124.738 \\
2MASXJ01144386 & 3.388 & 1.692 & 2.822 & 1.638 & 13.274 \\
2MASXJ08374557 & 10.715 & 1.649 & 3.722 & 1.596 & 16.866 \\
2MASXJ15570268 & 0.741 & 0.010 & 0.367 & 0.211 & 10.864 \\
2MASXJ16152860 & 0.724 & 0.108 & 0.981 & 0.538 & 0.551 \\
2MASXJ16301506 & 27.542 & 1.632 & 3.108 & 2.095 & 19.861 \\
2MASXJ17100856 & 0.020 & 0.034 & 0.268 & 0.245 & 1.782 \\
2MASXJ23024668 & 0.000 & 0.263 & 0.105 & 0.075 & 4.634 \\
ARP143 & 4.898 & 1.482 & 5.137 & 3.094 & 5.383 \\
ARP70 & 26.303 & 1.215 & 4.171 & 2.908 & 7.194 \\
CGCG004-035 & 3.890 & 0.542 & 2.596 & 1.455 & 5.012 \\
CGCG008-023 & 1.318 & 2.253 & 6.753 & 3.583 & 1.315 \\
CGCG047-117 & 64.565 & 3.356 & 5.592 & 3.931 & 14.454 \\
CGCG107-031 & 0.380 & 0.029 & 0.415 & 0.352 & 0.000 \\
CGCG207-042 & 3.467 & 1.071 & 3.151 & 2.100 & 40.832 \\
CGCG308-009 & 0.209 & 0.023 & 0.185 & 0.144 & 0.000 \\
CGCG476117 & 51.286 & 1.193 & 2.242 & 1.362 & 8.433 \\
FGC175A & 0.155 & 0.260 & 1.076 & 1.318 & 1.239 \\
GALEXASCJ012052 & 1.202 & 0.176 & 1.806 & 1.020 & 1.622 \\
GALEXASCJ234457 & 30.903 & 0.495 & 3.326 & 2.141 & 26.363 \\
IC0208 & 2.291 & 0.245 & 1.071 & 0.623 & 6.180 \\
IC0701 & 11.749 & 2.350 & 2.752 & 1.749 & 8.610 \\
IC0758 & 4.677 & 0.044 & 0.160 & 0.116 & 6.353 \\
IC1481 & 3.020 & 1.411 & 6.567 & 2.494 & 19.588
\enddata
\end{deluxetable}

\begin{deluxetable}{llrrrrrrrrrl}
\tablecaption{Properties of SNe Ia considered in this work. Light-curve fit parameters $x_0$, $x_1$ and $c$ were determined using the SALT2 supernova model. The first 25 entries are shown here. A full version of this table is available online.} \rotate \label{table:sn_properties}
\tablehead{\colhead{SN} & \colhead{Survey} & \colhead{$\mu$} & \colhead{RA} & \colhead{Dec} & \colhead{z} & \colhead{$x_0$} & \colhead{$x_1$} & \colhead{$c$} & \colhead{M$_\text{B}$} & \colhead{$\mu_{SN} - \mu_{mod}$} & \colhead{Galaxy}}
\startdata
ASASSN-14kd & CSPII & 35.001 & 343.353 & 4.799 & 0.023 & 0.006 & 1.090 & 0.144 & 16.207 & 0.025 & 2MASXJ22532475 \\
ASASSN-14lo & CSPII & 34.700 & 177.969 & 18.546 & 0.020 & 0.007 & -0.487 & 0.093 & 16.032 & 0.012 & UGC6837 \\
ASASSN-14my & CSPII & 34.706 & 174.626 & -8.976 & 0.020 & 0.011 & -0.578 & -0.060 & 15.550 & -0.045 & NGC3774 \\
ASASSN-15cb & CSPII & 36.204 & 189.959 & 3.797 & 0.040 & 0.003 & 0.918 & 0.025 & 17.049 & -0.031 & CGCG042-196 \\
ASASSN-15cd & CSPII & 36.117 & 149.812 & 12.988 & 0.034 & 0.004 & 0.322 & -0.097 & 16.670 & 0.226 & CGCG064-017 \\
ASASSN-15db & CSPII & 33.475 & 236.745 & 17.884 & 0.011 & 0.024 & -0.530 & 0.051 & 14.677 & 0.093 & NGC5996 \\
ASASSN-15dd & CSPII & 34.969 & 235.995 & 19.212 & 0.024 & 0.007 & -1.314 & -0.020 & 16.083 & -0.163 & CGCG107-031 \\
ASASSN-15ga & PS1 & 32.511 & 194.864 & 14.171 & 0.007 & 0.022 & -1.015 & 0.305 & 14.642 & 0.234 & NGC4866 \\
ASASSN-15tz & PS1 & 34.991 & 40.287 & 43.680 & 0.024 & 0.004 & -1.620 & 0.057 & 16.418 & -0.110 & UGC2164 \\
CSP13n & CSPII & 36.367 & 144.967 & 16.918 & 0.048 & 0.003 & 0.607 & -0.051 & 17.019 & -0.264 & CGCG092-024 \\
iPTF13dkl & PS1 & 36.759 & 356.242 & 3.395 & 0.040 & 0.001 & 0.631 & 0.001 & 17.580 & 0.525 & GALEXASCJ234457 \\
iPTF13dym & CSPII & 35.846 & 351.126 & 14.651 & 0.042 & 0.002 & -3.170 & 0.075 & 17.621 & -0.504 & 2MASSJ23243021 \\
iPTF14aaf & CSPII & 37.015 & 219.211 & 6.139 & 0.059 & 0.001 & 0.524 & 0.058 & 18.043 & -0.086 & CGCG047-117 \\
iPTF14bdn & MISC & 34.334 & 202.687 & 32.761 & 0.016 & 0.021 & 1.534 & -0.083 & 14.705 & 0.085 & UGC08503 \\
iPTF14gnl & CSPII & 36.896 & 5.951 & -3.856 & 0.054 & 0.002 & 0.587 & -0.081 & 17.453 & 0.002 & 2MASXJ00234829 \\
LSQ12gef & CSPII & 37.232 & 25.141 & 18.511 & 0.064 & 0.000 & 1.058 & 0.293 & 18.937 & -0.065 & 2MASXJ01403375 \\
LSQ12hzj & CSPII & 35.821 & 149.801 & -9.003 & 0.035 & 0.004 & -0.310 & -0.091 & 16.512 & -0.089 & 2MASXJ09591230 \\
LSQ13vy & CSPII & 36.243 & 241.732 & 3.001 & 0.042 & 0.002 & -1.105 & 0.087 & 17.671 & -0.088 & 2MASXJ16065563 \\
LSQ14ajn & CSPII & 34.918 & 178.877 & 11.924 & 0.021 & 0.006 & -1.950 & 0.017 & 16.273 & 0.113 & PGC37426 \\
LSQ14azy & CSPII & 36.601 & 168.145 & 12.072 & 0.046 & 0.001 & 0.182 & 0.073 & 17.742 & 0.065 & 2MASXJ11123493 \\
LSQ14q & CSPII & 37.456 & 133.488 & 17.328 & 0.067 & 0.001 & -0.430 & -0.094 & 18.160 & 0.062 & SDSSJ08535700 \\
LSQ15aae & CSPII & 36.755 & 247.563 & 5.931 & 0.052 & 0.001 & 1.120 & 0.127 & 17.899 & -0.049 & 2MASXJ16301506 \\
LSQ15agh & CSPII & 37.180 & 163.226 & 23.598 & 0.060 & 0.001 & 0.958 & -0.053 & 17.760 & 0.036 & 2MASXJ10525434 \\
PS15sv & CSPII & 35.651 & 243.299 & 1.592 & 0.033 & 0.005 & 0.319 & -0.031 & 16.423 & -0.173 & GALEXASCJ161311 \\
SN1989a & MISC & 32.926 & 171.992 & 29.510 & 0.009 & 0.046 & -0.104 & -0.010 & 13.848 & -0.103 & NGC3687
\enddata
\end{deluxetable}
 
\begin{deluxetable}{lrrrccccc}
\tablecaption{Galaxy coverage and fit successes. All photometry mass fits were successful.The first 25 entries are shown here. A full version of this table is available online.} \rotate \label{table:coverage}
\tablehead{\colhead{Galaxy} & \colhead{RA} & \colhead{Dec} & \colhead{$z$} & \colhead{GALEX} & \colhead{SDSS} & \colhead{2MASS} & \colhead{SFR} & \colhead{STARLIGHT Mass}}
\startdata
2MASSJ00164451 & 4.185 & 15.342 & 0.070 & X & X & X &  & X \\
2MASSJ00235669 & 5.986 & 29.522 & 0.073 & X & X & X &  & X \\
2MASSJ23243021 & 351.126 & 14.651 & 0.042 & X & X & X &  & X \\
2MASXJ22532475 & 343.353 & 4.799 & 0.024 & X & X & X & X & X \\
2MASXJ00234829 & 5.951 & -3.856 & 0.054 & X & X & X & X & X \\
2MASXJ01144386 & 18.683 & 0.286 & 0.045 & X & X & X & X & X \\
2MASXJ01403375 & 25.141 & 18.511 & 0.064 & X & X & X &  & X \\
2MASXJ02305208 & 37.717 & 22.476 & 0.031 & X & X & X &  & X \\
2MASXJ04424248 & 70.677 & 18.583 & 0.016 & X & X & X &  & X \\
2MASXJ07192718 & 109.863 & 54.229 & 0.067 & X & X & X &  & X \\
2MASXJ08374557 & 129.440 & 49.478 & 0.052 & X & X & X & X & X \\
2MASXJ09591230 & 149.801 & -9.003 & 0.033 & X & X & X &  & X \\
2MASXJ10525434 & 163.226 & 23.598 & 0.060 &  & X & X &  & X \\
2MASXJ11123493 & 168.145 & 12.072 & 0.046 & X & X & X &  &  \\
2MASXJ11500404 & 177.517 & 21.280 & 0.026 & X & X & X &  &  \\
2MASXJ12095669 & 182.486 & 47.096 & 0.031 &  & X & X &  & X \\
2MASXJ12385810 & 189.742 & 11.131 & 0.060 & X & X & X &  & X \\
2MASXJ15024995 & 225.708 & 48.784 & 0.026 &  & X & X &  & X \\
2MASXJ15393305 & 234.888 & 32.094 & 0.054 & X & X & X &  & X \\
2MASXJ15570268 & 239.261 & 37.417 & 0.031 & X & X & X & X & X \\
2MASXJ16065563 & 241.732 & 3.001 & 0.042 &  & X & X &  & X \\
2MASXJ16152860 & 243.869 & 19.226 & 0.033 & X & X & X & X & X \\
2MASXJ16301506 & 247.563 & 5.931 & 0.052 & X & X & X & X & X \\
2MASXJ17100856 & 257.536 & 74.729 & 0.042 & X & X & X & X & X \\
2MASXJ18242915 & 276.122 & 52.444 & 0.030 & X & X & X &  & X
\enddata
\end{deluxetable}

\begin{deluxetable}{lrrrrrrrrrrrrrr}
\tablecaption{In-house global photometry for SDSS and GALEX.  Only GALEX photometry with SDSS coverage was measured. Given as AB magnitudes; -99 marks negative flux. The first 25 entries are shown here. A full version of this table is available online.} \rotate \label{table:global_photometry}
\tablehead{\colhead{Galaxy} & \colhead{$m_u$} & \colhead{$\sigma_{m_u}$} & \colhead{$m_g$} & \colhead{$\sigma_{m_g}$} & \colhead{$m_r$} & \colhead{$\sigma_{m_r}$} & \colhead{$m_i$} & \colhead{$\sigma_{m_i}$} & \colhead{$m_z$} & \colhead{$\sigma_{m_z}$} & \colhead{$m_{NUV}$} & \colhead{$\sigma_{m_{NUV}}$} & \colhead{$m_{FUV}$} & \colhead{$\sigma_{m_{FUV}}$}}
\startdata
2MASSJ00164451 & 19.522 & 0.258 & 17.730 & 0.020 & 16.919 & 0.014 & 16.485 & 0.015 & 16.128 & 0.050 & -99.000 & -99.000 & -99.000 & -99.000 \\
2MASSJ00235669 & 17.894 & 0.054 & 16.737 & 0.005 & 16.492 & 0.007 & 16.280 & 0.009 & 16.109 & 0.028 & 18.676 & 0.048 & 19.239 & 0.106 \\
2MASSJ23243021 & 19.010 & 0.118 & 17.363 & 0.009 & 16.539 & 0.006 & 16.119 & 0.007 & 15.930 & 0.028 & 22.980 & 0.398 & 23.196 & 0.514 \\
2MASXJ00234829 & 16.298 & 0.046 & 15.597 & 0.006 & 15.074 & 0.007 & 14.675 & 0.008 & 14.783 & 0.030 & 17.661 & 0.034 & 18.158 & 0.064 \\
2MASXJ01144386 & 16.983 & 0.071 & 15.902 & 0.008 & 15.286 & 0.007 & 14.940 & 0.008 & 14.734 & 0.034 & 18.174 & 0.056 & 18.523 & 0.092 \\
2MASXJ01403375 & 17.235 & 0.092 & 15.842 & 0.008 & 15.147 & 0.007 & 14.735 & 0.007 & 14.384 & 0.018 & 18.693 & 0.090 & 18.925 & 0.135 \\
2MASXJ02305208 & 17.268 & 0.066 & 16.119 & 0.007 & 15.248 & 0.004 & 14.721 & 0.004 & 14.591 & 0.012 & 19.920 & 0.489 & -99.000 & -99.000 \\
2MASXJ08374557 & 17.472 & 0.077 & 16.399 & 0.011 & 15.936 & 0.011 & 15.509 & 0.011 & 15.342 & 0.033 & 18.813 & 0.097 & 19.552 & 0.212 \\
2MASXJ10525434 & 17.080 & 0.053 & 16.179 & 0.007 & 15.638 & 0.007 & 15.341 & 0.008 & 15.151 & 0.034 & -99.000 & -99.000 & -99.000 & -99.000 \\
2MASXJ11123493 & 17.033 & 0.052 & 16.064 & 0.007 & 15.447 & 0.006 & 15.068 & 0.007 & 14.688 & 0.025 & 18.422 & 0.050 & 19.284 & 0.163 \\
2MASXJ11500404 & 17.178 & 0.039 & 15.635 & 0.005 & 14.875 & 0.003 & 14.513 & 0.004 & 14.262 & 0.009 & 20.588 & 0.222 & -99.000 & -99.000 \\
2MASXJ12095669 & 17.249 & 0.048 & 15.797 & 0.005 & 14.970 & 0.004 & 14.565 & 0.004 & 14.281 & 0.012 & -99.000 & -99.000 & -99.000 & -99.000 \\
2MASXJ12385810 & 17.352 & 0.079 & 16.419 & 0.011 & 15.782 & 0.010 & 15.459 & 0.011 & 15.080 & 0.028 & 19.298 & 0.116 & 19.734 & 0.248 \\
2MASXJ15024995 & 17.948 & 0.092 & 16.413 & 0.012 & 15.845 & 0.009 & 15.433 & 0.010 & 15.326 & 0.030 & -99.000 & -99.000 & -99.000 & -99.000 \\
2MASXJ15393305 & 16.579 & 0.054 & 15.286 & 0.005 & 14.449 & 0.004 & 14.054 & 0.004 & 13.822 & 0.011 & 18.930 & 0.101 & 22.090 & 1.304 \\
2MASXJ15570268 & 17.752 & 0.083 & 16.374 & 0.008 & 15.580 & 0.006 & 15.280 & 0.007 & 14.869 & 0.025 & 20.112 & 0.295 & 20.519 & 0.546 \\
2MASXJ16065563 & 16.311 & 0.056 & 15.230 & 0.008 & 14.891 & 0.008 & 14.522 & 0.008 & 14.502 & 0.036 & 17.177 & 0.037 & 17.599 & 0.107 \\
2MASXJ16152860 & 18.085 & 0.134 & 16.823 & 0.012 & 16.104 & 0.010 & 15.700 & 0.012 & 15.442 & 0.032 & 20.148 & 0.189 & 21.911 & 1.727 \\
2MASXJ16301506 & 16.960 & 0.064 & 15.789 & 0.008 & 15.144 & 0.006 & 14.761 & 0.006 & 14.468 & 0.022 & 18.730 & 0.105 & 19.668 & 0.294 \\
2MASXJ17100856 & 17.798 & 0.076 & 16.191 & 0.006 & 15.471 & 0.005 & 15.108 & 0.005 & 14.809 & 0.019 & 21.724 & 0.668 & 20.789 & 0.429 \\
2MASXJ21352164 & 17.328 & 0.130 & 15.943 & 0.009 & 15.062 & 0.006 & 14.499 & 0.005 & 14.250 & 0.015 & -99.000 & -99.000 & 20.613 & 1.575 \\
2MASXJ22532475 & 17.066 & 0.077 & 15.499 & 0.005 & 14.814 & 0.005 & 14.483 & 0.005 & 14.212 & 0.020 & 18.084 & 0.059 & 19.099 & 0.163 \\
2MASXJ23024668 & 17.992 & 0.104 & 16.219 & 0.006 & 15.345 & 0.004 & 14.902 & 0.004 & 14.561 & 0.010 & 21.219 & 0.097 & 22.513 & 0.326 \\
ARP143 & 13.572 & 0.007 & 12.328 & 0.001 & 11.720 & 0.001 & 11.417 & 0.001 & 11.179 & 0.004 & 14.712 & 0.010 & 15.149 & 0.019 \\
ARP70 & 15.566 & 0.047 & 14.259 & 0.004 & 13.531 & 0.003 & 13.068 & 0.003 & 12.797 & 0.010 & 17.264 & 0.033 & 18.020 & 0.075
\enddata
\end{deluxetable}

\begin{deluxetable}{lrrrrrr}
\tablecaption{In-house global photometry 2MASS.  Only 2MASS photometry with SDSS coverage was measured. Given as AB magnitudes. -99 marks negative flux. The first 25 entries are shown here. A full version of this table is available online.} \rotate \label{table:global_photometry_2mass}
\tablehead{\colhead{Galaxy} & \colhead{$m_H$} & \colhead{$\sigma_{m_H}$} & \colhead{$m_J$} & \colhead{$\sigma_{m_J}$} & \colhead{$m_K$} & \colhead{$\sigma_{m_K}$}}
\startdata
2MASSJ00164451 & 15.485 & 0.397 & 16.047 & 0.451 & 16.439 & 1.304 \\
2MASSJ00235669 & 15.755 & 0.479 & 16.203 & 0.475 & 18.210 & 5.562 \\
2MASSJ23243021 & 15.358 & 0.324 & 15.465 & 0.268 & 15.180 & 0.404 \\
2MASXJ00234829 & 13.329 & 0.171 & 14.607 & 0.326 & 14.105 & 0.417 \\
2MASXJ01144386 & -99.000 & -99.000 & 14.672 & 0.377 & 14.144 & 0.455 \\
2MASXJ01403375 & 13.856 & 0.208 & 14.157 & 0.182 & 13.793 & 0.242 \\
2MASXJ02305208 & 13.160 & 0.100 & 13.805 & 0.104 & 13.574 & 0.134 \\
2MASXJ08374557 & 14.847 & 0.462 & 15.332 & 0.506 & 14.678 & 0.456 \\
2MASXJ10525434 & -99.000 & -99.000 & 15.940 & 0.948 & -99.000 & -99.000 \\
2MASXJ11123493 & 14.657 & 0.504 & 14.719 & 0.399 & 14.487 & 0.567 \\
2MASXJ11500404 & 12.939 & 0.120 & 13.699 & 0.097 & 13.465 & 0.150 \\
2MASXJ12095669 & 14.019 & 0.268 & 14.034 & 0.156 & 14.034 & 0.250 \\
2MASXJ12385810 & 16.121 & 2.423 & 15.263 & 0.731 & 14.925 & 0.922 \\
2MASXJ15024995 & 14.778 & 0.570 & 15.113 & 0.463 & 14.607 & 0.453 \\
2MASXJ15393305 & 12.818 & 0.127 & 13.347 & 0.129 & 13.106 & 0.165 \\
2MASXJ15570268 & 14.630 & 0.406 & 14.387 & 0.203 & 14.981 & 0.547 \\
2MASXJ16065563 & 13.338 & 0.244 & 13.982 & 0.316 & 13.817 & 0.426 \\
2MASXJ16152860 & 13.842 & 0.192 & 14.678 & 0.251 & 14.398 & 0.321 \\
2MASXJ16301506 & 13.701 & 0.179 & 13.994 & 0.158 & 13.769 & 0.232 \\
2MASXJ17100856 & 14.163 & 0.183 & 14.317 & 0.139 & 13.955 & 0.189 \\
2MASXJ21352164 & 13.649 & 0.231 & 13.797 & 0.180 & 14.021 & 0.288 \\
2MASXJ22532475 & 14.868 & 0.777 & 13.796 & 0.187 & 14.147 & 0.419 \\
2MASXJ23024668 & 13.672 & 0.143 & 14.053 & 0.097 & 14.050 & 0.158 \\
ARP143 & 11.349 & 0.082 & 10.893 & 0.035 & 11.070 & 0.066 \\
ARP70 & 12.872 & 0.270 & 12.415 & 0.098 & 12.783 & 0.194
\enddata
\end{deluxetable}

\bibliographystyle{aasjournal}
\bibliography{main}

\end{document}